\documentclass{article}
\usepackage{amsmath,subfigure,chemarr}
\usepackage{amssymb}
\usepackage{graphicx}

\newcommand{\p}{\partial}
\newcommand{\ds}{\displaystyle}
\newcommand{\beq}{\begin{eqnarray}}
\newcommand{\beqq}{\begin{eqnarray*}}
\newcommand{\eeq}{\end{eqnarray}}
\newcommand{\eeqq}{\end{eqnarray*}}
\newcommand{\eps}{\varepsilon}

\begin{document}
\title{Stochastic modeling of gene activation and application to cell regulation}
\author{G. Malherbe David Holcman \footnote{Department of Computational Biology, IBENS,
 Ecole Normale Sup\'erieure, 46 rue d'Ulm 75005 Paris,
France. This research is supported by an ERC Starting Grant.}}
\maketitle{}

\begin{abstract}
Transcription factors (TFs) are key regulators of gene expression.
Based on the classical scenario in which the TF search process
switches between one-dimensional motion along the DNA molecule and
free Brownian motion in the nucleus, we study the arrival time of
several TFs to multiple binding sites and derive, in the presence of
competitive binding ligands, the probability that several target
sites are bound. We then apply our results to the hunchback
regulation by bicoid in the fly embryo and we propose a general mechanism that
allows cells to read a morphogenetic gradient and specialize
according to their position in the embryo.
\end{abstract}

\textbf{Keywords:}
modeling, stochastic binding, diffusion of transcription factor,
gene activation, morphogenetic gradient, cell differentiation.

\section{Introduction}
Transcription factors (TFs) are key regulators that can initiate or
inhibit gene activation by binding to specific DNA sites. TFs enter
the cell nucleus and search for their specific binding sites on the
DNA molecule. In a context of competition for activation and
inhibition, the search for target sites should not be too long
otherwise, another gene might be activated. This is the case for
olfactory gene activation \cite{Firestein}, where a single G-coupled
receptor out of hundreds is expressed in a single cell, while all
other receptors are repressed. Thus cells might use various
mechanisms to control gene activation including changing local
properties of the DNA molecule (for example by methylation or
acetylation) or changing the properties of the TF interaction as
what happens when specific molecules bind to TFs and modify their
affinity for the DNA molecule.

The mean time to reach a target site is thus a fundamental parameter
of gene activation and several biophysical scenarios have been
proposed to estimate it. Berg and Von Hippel \cite{VonHippel,Berg,blomberg} realized that the search time cannot
be computed using a three dimensional random walk only, because the
observed search time is indeed shorter. Using the property that the
TF scans the DNA base pairs for potential binding sites, they
proposed that the search consists of consecutive cycles of three dimensional free diffusion in the nucleus and
one dimensional random motion along the DNA molecule. DNA base pairs
are not electrically charged and so no long distance interactions
are involved, thus, the TF should physically come close to the DNA
in order to generate a true interaction. During the one dimensional
search, the TF is confined to a neighborhood of the DNA molecule and
can detach due to thermal fluctuations and a new 1d-3d cycle resumes
until the target is eventually reached. This basic scenario has been
confirmed experimentally by single particle tracking experiments
\cite{Elf} and investigated theoretically, by accounting for the
local base pair interactions, the mean number to scan the base pairs
per cycle, the free diffusion time and the time the TF is bound to
the DNA molecule \cite{Slutsky,Wunderlich,Halford,Hu,PLA}.

However, it is still unclear how to go from the TF search time to
the mechanism responsible for cell specialization. Cells in a living
tissue are imbedded in a matrix of positional information, generated
by morphogenetic gradients \cite{turing,wolpert,meinhart,monk1,monk2}. A first step consists of
the ability of the cell to "read out" the local characteristics of
the morphogenetic gradient so that the cell can be labeled and
acquire its own identity. To address this question, here we
investigate how several binding sites can be bound at steady state
by several TFs generated by an external steady state gradient. In
particular, we are led to compute the mean time for some TFs to bind
to many target sites.  The number of bound sites can be seen as a
digital-converter used by the cell to discriminate two different
morphogenetic gradient concentrations depending whether or not all
the binding sites are saturated. We further study the effect of
competitive ligands that can modulate
the TF's action leading to gene repression.

First, we present our computation for the time for a single TF to
bind to one of multiple potential binding sites. The distribution of the arrival
time is always a sum of two exponentials but, for a large number of
free diffusion and DNA binding cycles, the arrival time is single
exponentially distributed. We then expand our analysis to multiple
TFs with multiple targets in the presence of enzymes leading to
degradation. We apply our results to estimate the number of active
sites when the cell nucleus experiences a steady state TF influx.
Finally, we estimate the steady state probability that a given
number of binding sites are occupied. In our model, this probability
describes the proportion of time a gene is actively transcribed. We
apply our analysis to the initial patterning of the fly embryo by
the bicoid (bcd) morphogenetic gradient. The bcd gradient regulates
a number of downstream TFs involved in the gap gene network
\cite{Feng,Ashyraliyev}, which determines the position of body
sections along the anterior-posterior (A-P) axis in the drosophila
embryo. Among these gap genes, hunchback (hb) is responsible for
thoracic development \cite{Feng,Ashyraliyev}. Hb activation leads to
the formation of a sharp boundary and to the formation of stripes.
We use our analysis of TF binding to determine the hb density
induced by bcd activation, and we show that this hb-bcd
interaction-modulation is sufficient to generate the transition from
a smooth bicoid gradient into a sharp hb boundary in the middle of
the drosophila embryo. Our approach provides a general scenario at a
molecular level of TF interactions that lead to cell specialization.

\section {Distribution and mean of the search time}
We first summarize the properties associated with the TF's search
process to its binding site. The TF switches between a free
diffusion and random walk along the DNA molecule
\cite{VonHippel,Slutsky,Benichou,PLA}.
\begin{enumerate}
\item Due to the interaction potential with the DNA backbone \cite{Nadassy},
the TF can bind unspecifically to the DNA molecule. The strength of
the interaction potential is around $16 kT$ \cite{O'Gorman,Slutsky},
larger than the thermal noise $\sim kT$. In this deep well
approximation, the random time $\tau_{d}$ a TF stays bound is
exponentially distributed \cite{SchussBook}. Experimental and
theoretical estimates for the average time $\overline{\tau}_{d}$ are
in the range of a few milliseconds \cite{Elf,PLA}.
\item A bound TF slides along the DNA molecule during a random time
$\tau_{d}$ scanning an average number $\overline{n}(\tau_{d})$ of
base pairs (bp). The mean number
$\overline{n}=\mathbb{E}_{\tau_{d}}(\overline{n}(\tau_{d}))$ of base
pairs scanned before detaching is on the order of 100
\cite{Elf,PLA}.
\item A TF can detach from the DNA due to thermal noise and diffuse freely
in the nucleus until it rebinds to the DNA. When the DNA molecule
occupies a small fraction of the nucleus and can be approximated as
long rods, the random time $\tau_{f}$ a TF spends diffusing in the
nucleus is exponentially distributed \cite{PLA} with an average
$\overline{\tau}_{f}$, which is on the order of a few milliseconds
\cite{Elf,PLA}. However, for larger density and a complex DNA
organization, the distribution time in general is a sum of
exponentials and might even be more complicated.
\end{enumerate}

We start with $n_f$ copies of a TF, alternating independently
between periods of free diffusion and random walks along the DNA
until one of the $n_s$ binding sites is found. We further consider
competitive ligands that can bind to the TF target sites, preventing
the sites to be occupied by TFs. The ligand $L$ binds to the target
site $S$ according to a first order reaction:
\beq
\label{drugreaction}
S+L \mathop{\rightleftharpoons}^{k_{a}}_{k_{d}}S.L,
\eeq
with an association and a dissociation rate $k_{a}$ and $k_{d}$
respectively. At equilibrium, Michaelis-Menten reaction says that
for a concentration $C$ of ligands, the probability that a binding
site is not occupied is:
\beq
P=\frac{1}{1+C\frac{k_{a}}{k_{d}}}.\label{pdexpression}
\eeq
%
\subsection*{Search time for a single TF}
The random time $T(1,n_s)$ for a single TF to bind to a target site
is the sum of the time spent in one and three dimensions. Using the
characteristic function of the search time $T(1,n_s)$, we can
express the probability density function (pdf)
\beq
p_T(t)=\frac{d}{dt}Pr\{T(1,n_s)<t\}
\eeq
as follows (see appendix):
\beq\label{eqpT}
p_T(t)=\frac{r_2}{r_2-r_1}\frac{e^{-r_1t}}{r_1}+\frac{r_1}{r_1-r_2}\frac{e^{-r_2t}}{r_2},
\eeq
where $r_1$ and $r_2$ are the two positive roots of
$\left(1-x\overline{\tau}_{d}\right)\left(1-x\overline{\tau}_{f}\right)-1+p(n_s)=0$
and $p(n_s)$ is the probability to find a target during a single one dimensional walk along the DNA. The associated mean binding time is:
\beq
\overline{T}(1,n_s)=\int\limits_0\limits^\infty
t\,p_T(t) dt=\frac{r_2}{r_1(r_2-r_1)}+\frac{r_1}{r_2(r_1-r_2)}.
\eeq
In the limit $p(n_s)\ll 1$, using the expression for the two roots
and approximating the pdf $p_T$ (eq.~\ref{eqpT}) by a single
exponential for a time t such that $
\left(\frac{1}{\overline{\tau}_{d}}+\frac{1}{\overline{\tau}_{f}}\right)t\gg
1$ (see appendix), we obtain that
\beq\label{pTsingleexp}
p_T(t) =
\frac{p(n_s)}{\overline{\tau}_{d}+\overline{\tau}_{f}}e^{-\frac{p(n_s)}{\overline{\tau}_{d}+\overline{\tau}_{f}}t}.
\eeq
Since $\overline{\tau}_{d}$ and $\overline{\tau}_{f}$ are both on
the order of a few ms \cite{PLA,Elf}, the single exponential limit
is valid for $t$ larger than a few ms. The mean time
$\overline{T}(1,n_{s})$ then reduces to
\beq
\overline{T}(1,n_{s})\approx\frac{\overline{\tau}_{d}+\overline{\tau}_{f}}{p(n_s)}.
\eeq
The mean number of free diffusions and DNA bindings before finding
the target site is equal to $\frac{1}{p(n_s)}$. The limit
$p(n_s)\ll1$ corresponds to TFs that find their target sites after a
large number of cycles.

\subsection*{Search time for multiple TFs}
When there are $n_f$ TFs that can potentially bind to $n_s$ identical
binding sites (a site can only be occupied by a single TF), in the single
exponential limit, the time $T(n_f,n_{s})$ for the first TF to bind
a site is the minimum of the $n_f$ exponential laws of mean time
$\overline{T}(1,n_{s})$. $T(n_f,n_{s})$ is then exponentially
distributed with mean:
\beq
\overline{T}(n_f,n_{s})=\frac{\overline{T}(1,n_{s})}{n_f}.
\eeq
We shall consider $n_s$ well separated sites (by at least a distance
of $\overline{n}$ base pairs). In this case, the probability to find
each site during a DNA biding is $\frac{\overline{n}}{N_{bp}}$ where
$N_{bp}$ is the total number of base pairs in the genome.
Furthermore, in the presence of competitive ligands, there are $P
n_s$ available binding sites. Thus, the probability of binding to
one of the $n_s$ sites is $p(n_s)=P
n_{s}\frac{\overline{n}}{N_{bp}}$ and the mean binding time for
$n_s$ well separated sites with $p(n_s)\ll1$ is:
\beq
 \overline{T}(n_f,n_s)
&\approx&\frac{\overline{\tau}_{d}+\overline{\tau}_{f}}{n_fp(n_s)}=\frac{(\overline{\tau}_{d}+\overline{\tau}_{f})N_{bp}}{n_fPn_{s}
\overline{n}}\\&=&\frac{\overline{T}_{S}}{n_fn_{s}},
 \label{tempsmultpileunsite1}
\eeq
where
\beq
\overline{T}_{S}=\frac{(\overline{\tau}_{d}+\overline{\tau}_{f})N_{bp}}{P\overline{n}}
\label{timefindtarget}
\eeq
is the search time for a single TF with a single target site.

{\noindent \bf Remark 1.} Formula (\ref{tempsmultpileunsite1})
describes the combined effect of multiple but well separated binding
sites. When the sites are clustered, the mean time to find a target
becomes a nonlinear function of the distribution
\cite{HS-hole1,HS-hole2,WardM1,WardM2,WardM3} and has been
approximated by the Berg-Purcell approximation formula
\cite{Berg-Purcell}. When there are $n_s$ binding sites of size $a$,
located on an ensemble of DNA-molecules on a sphere of radius R, the
mean time $\tau_d$ in 3d to find a site is:
\beq
\tau_d\approx\frac{|\Omega|}{D_H}\left( \frac{1}{4\pi R}+\frac{1}{4 n_s a}\right).
\eeq
This formula can be improved \cite{zwanzig,bere}. Here $D_H$ is an
effective diffusion constant that accounts for the switch between
the 1D DNA motion and the 3D diffusion. When the one 1D excursion
length is small compared to the 3D diffusion length,
\beq
D_H\approx\ds {\frac{D}{1+\frac{\tau_{d}}{\tau_f}}}.
\eeq
In the other cases, one has to deal with random jumps.

{\noindent \bf Remark 2}  For $P=1$ (no competitive ligand),
$\overline{n} \approx 100$ \cite{Elf,Slutsky,PLA} and for a
relatively small genome $N_{bp}=10^6$, $p(n_s)$ is approximated by:
\beq
p(n_s)\approx n_s 10^{-4}.
\eeq
Thus $p(n_s)\ll 1$ is valid as long as the number of binding sites
satisfies $n_s\ll 10^4$. We conclude that $p(n_s)\ll1$ is verified
in most cases.

\section {From a morphogenetic gradient to DNA site activation}
We shall now apply our previous results to estimate the number of
occupied sites when a nucleus receives a steady influx of TFs. This
steady influx of TFs entering the nucleus could, for example, either
be imported from outside the cell or be steadily produced in the
cytoplasm of the cell. We consider that gene expression is
proportional to the mean time the binding sites are occupied. Since
the number of binding sites occupied controls gene expression, we
shall estimate, for a given TF influx, the mean proportion of time
the binding sites are occupied.

\subsection{Activation of a single binding site}
\begin{figure}
\centerline{\subfigure[$\mathbb{P}_1$ as a function of $\alpha$ ]{\includegraphics[width=6cm]{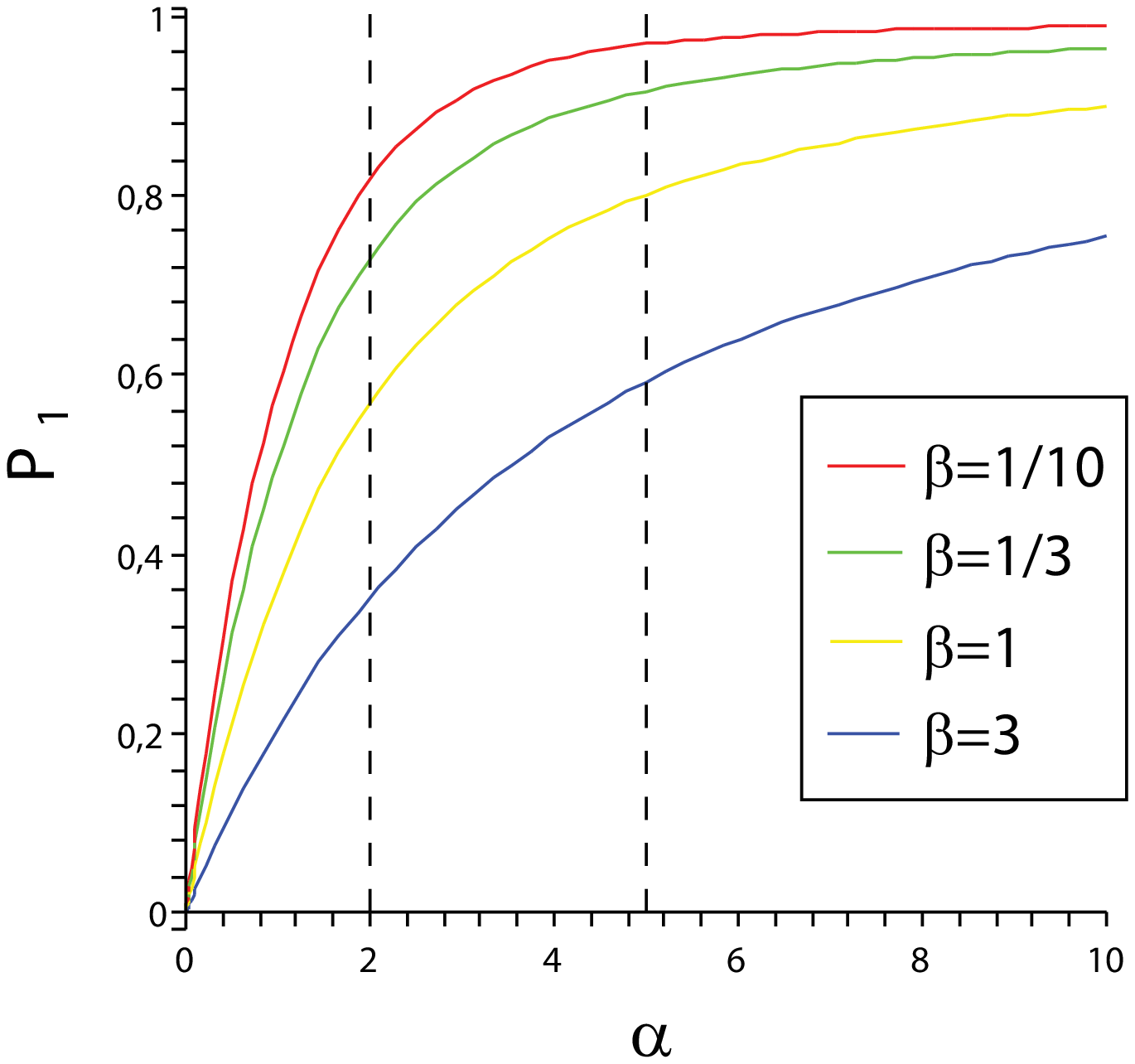}}\subfigure[ $\mathbb{P}_1$ as a function of the competitor concentration $C$]{
\includegraphics[width=6cm]{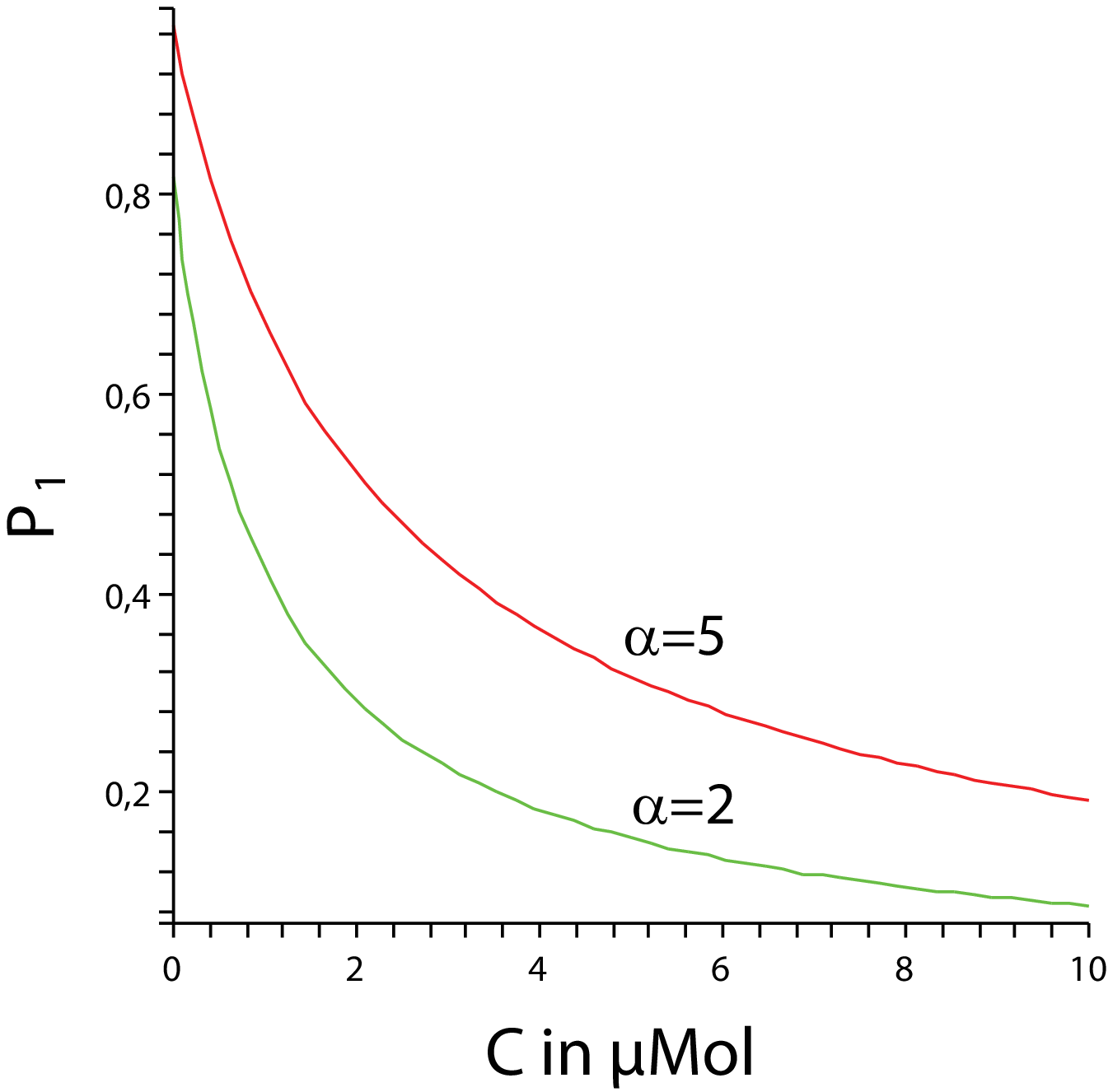}}}
\caption[(a) \textit{$\mathbb{P}_1$ as a function of $\alpha$}\newline(b) \textit{$\mathbb{P}_1$ as a function of the competitor concentration $C$}]{\textbf{(a) $\mathbb{P}_1$ as a function of $\alpha$ for various values of $\beta$.} From left to right, $\beta$ increases $1/10$ (red),$1/3$ (green),1 (yellow),3 (blue). The upper curves correspond to fast search times and/or long binding times to the target site and no competitors.  \textbf{(b) $\mathbb{P}_1$ as a function of the competitor concentration $C$ in $\mu$Mol.} The upper curve is obtained for $\alpha=5$, the lower one is for $\alpha=2$, where $\beta=\beta_0\left(1+C\frac{k_{a}}{k_{d}}\right)$ with $\beta_0=\frac{1}{10}$ for $C=0$ and $\frac{k_{a}}{k_{d}}=20\mu$Mol$^{-1}$ \cite{Baliga}.}
\label{reponse1site}
\end{figure}

We first compute the average occupation ratio $\mathbb{P}_1$ of a
single binding site before considering multiple sites in the
following section. To compute  $\mathbb{P}_1$, we use Bayes' law and
sum over the number of TFs in the nucleus:
\beq
\mathbb{P}_1= \sum\limits_{n_f=0}\limits^{+\infty}\mathbb{P}(1|n_f)\mathbb{P}(n_f)\label{decompose1site},
\eeq
where $\mathbb{P}(n_f)$ is the probability of having $n_f$ TFs in
the nucleus and $\mathbb{P}(1|n_f)$ is the conditional probability
that a single binding site is occupied when there are $n_f$ TFs. To
proceed with the computation of $\mathbb{P}_1$, we assume that TFs
arrive in the nucleus at a Poissonnian rate $\lambda$ and are
degraded (free or bound) by enzymes at a rate $K$. Thus, the number
of TFs in the nucleus follows a birth and death process and is
distributed according to a Poisson law with mean
$\alpha=\frac{\lambda}{K}$:
\beq
\mathbb{P}(n_f)=\frac{\alpha^{n_f}}{n_f!}e^{-\alpha}.\label{probantf}
\eeq
We now compute $\mathbb{P}(1|n_f)$.  When a TF has found the target,
it stays attached for a mean time $\overline{T}_b$. We consider that
the rate of binding and unbinding to the sites is faster than the
rate of TF turn over in the nucleus and that the steady state
between binding and unbinding is reached, thus
\beq
\mathbb{P}(1|n_f)=\frac{\overline{T}_{b}}{\overline{T}_{b}+\overline{T}(n_f,1)}=\frac{\overline{T}_{b}}{\overline{T}_{b}+\frac{\overline{T}_{S}}{ n_{TF}}}=\frac{n_f}{n_f+\beta}, \label{conditional1site}
\eeq
where $\beta=\frac{\overline{T}_{S}}{\overline{T}_b}.$ Using equations
(\ref{decompose1site}), \ref{probantf} and
(\ref{conditional1site}),  we get:
\beq
\mathbb{P}_1&=&e^{-\alpha}\sum\limits_{n_f=1}\limits^{\infty}\frac{n_f}{n_f+\beta}
\frac{\alpha^{n_f}}{n_f!}. \label{equation1}
\eeq
Using
$\frac{\alpha^{n_f}}{n_f+\beta}=\alpha^{-\beta}\int\limits_0\limits^{\alpha}x^{\beta+n_f-1}dx,$
we obtain:
\beq
\mathbb{P}_1&=&e^{-\alpha}\alpha^{-\beta}
\sum\limits_{n_f=1}\limits^{+\infty}\int\limits_0\limits^{\alpha}\frac{x^{\beta+n_f-1}}{(n_f-1)!}
dx\nonumber\\
&=&e^{-\alpha}\alpha^{-\beta}
\int\limits_0\limits^{\alpha}x^{\beta} \left(\sum\limits_{n_f=1}\limits^{+\infty}
x^{n_f-1}\frac{1}{(n_f-1)!}\right) dx\nonumber\\ &=&e^{-\alpha}\alpha^{-\beta}
\int\limits_0\limits^{\alpha}x^{\beta} e^x dx\nonumber\\&=& \alpha\int\limits_0\limits^{1}u^{\beta}
e^{\alpha(u-1)} du,\label{integ1}
\eeq
where $x=\alpha u$. We plot in figure
\ref{reponse1site}a the occupation ratio $\mathbb{P}_1$ as
a function of the average number $\alpha$ of TFs in the nucleus for
different values of $\beta$. When the competitor ligand
concentration $C$ varies, the occupation ratio is modulated as
described in figure \ref{reponse1site}b.  Using Lac I data
\cite{Elf} and in the absence of DNA binding competitor ($P=1$), the
total search time is $\overline{T}_{S}=6\min$
\cite{Elf,PLA}, while $\overline{T}_{b}\approx 70 \min$
\cite{Lin} and thus the ratio is $\beta=\frac{\overline{T}_{S}}{\overline{T}_{b}} \approx1/11$.
We conclude (red curve Fig \ref{reponse1site}a) that for low
$\beta$, the target site can be occupied for a significant
proportion of time. In particular, small fluxes of TFs can induce
significant modifications on gene expression in a target cell.


\subsection{Activation with multiple binding sites}
When there are $n_{s}$ binding sites, we shall now compute the
proportion of time $\mathbb{P}_k $ that $k$ sites are occupied.
Using Bayes' law, we have:
\beq
\mathbb{P}_k = \sum\limits_{n_f=0}^{\infty}\mathbb{P}(k |n_f)\mathbb{P}(n_f),
\label{decompnsites}
\eeq
where $\mathbb{P}(n_f)$ is the probability to have $n_f$ TFs given
by expression (\ref{probantf}). We now compute $\mathbb{P}(k |n_f)$
by analyzing a Markov chain \cite{HolcmanJCP2005} which
describes the probability $\mathbb{P}_q(t)$ that q sites are
occupied at time $t$.

When $q$ sites are occupied, the total release rate is
$\frac{q}{\overline{T}_{b}}$ while the arrival rate to a site is
given by
$\overline{T}^{-1}(n_f-q,n_s-q)=\frac{(n_f-q)(n_{s}-q)}{\overline{T}_{S}}$
with equation (\ref{timefindtarget}). The forward and backward rate
of the Markov chain are given by:
\beq
F_q&=&\frac{(n_f-q)(n_{s}-q)}{\overline{T}_{S}}\\
B_q&=&\frac{q}{\overline{T}_{b}},
\eeq
and the Markov chain is given by \cite{HolcmanJCP2005}:
\beq
\frac{d}{dt}\mathbb{P}(q,t|n_f)=
-(F_q+B_q)\mathbb{P}(q,t|n_f)+F_{q-1}
\mathbb{P}(q-1,t|n_f)+B_{q+1}\mathbb{P}(q+1,t|n_f),
\eeq
with the boundary conditions:
\beq
\frac{d}{dt}\mathbb{P}(n_f,t|n_f)&=&
F_{n_f-1}
\mathbb{P}(n_f-1,t|n_f) -B_{n_f}\mathbb{P}(n_f,t|n_f)\\
\frac{d}{dt}\mathbb{P}(0,t|n_f)&=&
-F_0\mathbb{P}(0,t|n_f)+B_{1}\mathbb{P}(1,t|n_f).
\eeq
We consider that the rate of binding and unbinding to the sites is
faster than the rate of TF turn over in the nucleus and that the
steady state is achieved quickly, thus:
\beq
0=
-(F_q+B_q)\mathbb{P}(q|n_f)+F_{q-1}
\mathbb{P}(q-1|n_f)+B_{q+1}\mathbb{P}(q+1|n_f),
\eeq
where $\mathbb{P}(q|n_f)=\mathbb{P}(q,\infty|n_f).$  By induction
\cite{HolcmanJCP2005}, for  $k\leq n'=\min(n_f,n_s)$ (the maximal
number of sites occupied by TFs), we get:
\beq
\mathbb{P}(k |n_f) = \mathbb{P}(0 |n_f)\frac{1}{\beta^{k}k!}\prod\limits_{j=0}\limits^{k-1}(n_f-j)(n_{s}-j)
\label{eigenvector},
\eeq
where:
\beq
\beta=\frac{\bar{T}_{S}}{\overline{T}_b}.
\eeq
For $k>n'=\min(n_f,n_s)$, $\mathbb{P}(k |n_f) =0$ since there can be no more than  $n'$ TFs bound. Using the normalization condition,
\beq \label{normilization}
\sum \limits_{k=0}\limits^{n'}\mathbb{P}(k |n_f)=1,
\eeq
we finally get for $1\leq k\leq n'$:
\beq
\mathbb{P}(k |n_f) = \frac{\frac{1}{\beta^{k}k!}\prod\limits_{j=0}\limits^{k-1}(n_f-j)(n_{s}-j)}
{1+\sum\limits_{l=1}\limits^{n'}
\frac{1}{\beta^{l}l!}\prod\limits_{j=0}\limits^{l-1}(n_f-j)(n_{s}-j)}.
\label{eigenvector2}
\eeq
and for $k=0$:
\beq
\mathbb{P}(0 |n_f) = \frac{1}
{1+\sum\limits_{l=1}\limits^{n'}
\frac{1}{\beta^{l}l!}\prod\limits_{j=0}\limits^{l-1}(n_f-j)(n_{s}-j)}.
\label{eigenvector2-0}
\eeq
Using expressions (\ref{probantf}), (\ref{decompnsites}) and
(\ref{eigenvector}), we obtain for $1\leq k\leq n_s$:
\beq
\mathbb{P}_k = \sum\limits_{n_f=k}\limits^{\infty}\frac{\alpha^{n_f}}{n_f!}e^{-\alpha}\frac{\frac{1}{\beta^{k}k!}\prod\limits_{j=0}\limits^{k-1}(n_f-j)(n_{s}-j)}
{1+\sum\limits_{l=1}\limits^{\min(n_f,n_s)}
\frac{1}{\beta^{l}l!}\prod\limits_{j=0}\limits^{l-1}(n_f-j)(n_{s}-j)},
\label{pkcomplete}
\eeq
and for $k=0$:
\beq
\mathbb{P}_0 =e^{-\alpha}+ \sum\limits_{n_f=1}\limits^{\infty}\frac{\alpha^{n_f}}{n_f!}e^{-\alpha}\frac{1}
{1+\sum\limits_{l=1}\limits^{\min(n_f,n_s)}
\frac{1}{\beta^{l}l!}\prod\limits_{j=0}\limits^{l-1}(n_f-j)(n_{s}-j)},
\label{pkcomplete2}
\eeq
and $\mathbb{P}_k=0$ for $k>n_s$ as there can not be more than $n_s$
TFs bound to the target sites.
We shall now derive asymptotic
expressions for $\mathbb{P}_k$ when $\alpha\ll 1$ and $\beta\ll 1$,
which correspond respectively to a small average number of TFs in
the nucleus and to TFs that stay bound to the targets a long time
compared to the search time.

\subsection*{Asymptotics for $\alpha$ small}
With the expression of $\mathbb{P}(n_f)$  given in (\ref{probantf}) and the summation (\ref{decompnsites}), only the terms $n_f=0,1$ contribute to the first order asymptotic in $\alpha\ll 1$. With (\ref{eigenvector}) we obtain:
\beq
\mathbb{P}(1|1)
=\frac{n_s}{\beta}\mathbb{P}(0|n1)
\eeq
and with $\mathbb{P}(1|1) +\mathbb{P}(0|1)\approx1$,
\beqq
\mathbb{P}(1|1) &=& \frac{n_s}{n_s+\beta}\nonumber\\
\mathbb{P}(0|1) &=&1-\mathbb{P}(1|1).\nonumber
\eeqq
With (\ref{decompnsites}), for $\alpha\ll 1$,
\beq
\mathbb{P}_{1}&\approx&\alpha e^{-\alpha}\frac{n_s}{n_s+\beta}
\approx\frac{\alpha n_s}{n_s+\beta}.
\label{betasmallmult}
\eeq
We conclude that the probability that one site is occupied is given by the average number of TFs $\alpha$ multiplied by the probability
$\frac{n_s}{n_s+\beta}$ to have one site occupied when there is one TF in the nucleus.

\subsection*{Asymptotic for $\beta$ small}
We now compute the asymptotic for $\beta\ll 1$.

\begin{enumerate}
\item When the number of TFs is larger than the number of available sites ($n_f\geq n_{s}$), using equation (\ref{eigenvector2}), for $\beta\ll 1$, only the terms $\mathbb{P}(n_s-1|n_f)$ and $\mathbb{P}(n_s |n_f)$ contribute to the first order for $\beta\ll1$. Using the normalization relation (\ref{normilization}), \beq \mathbb{P}(n_s-1|n_f)+\mathbb{P}(n_s |n_f)\approx 1.\nonumber \eeq
    Furthermore with (\ref{eigenvector2}), \beq \mathbb{P}(n_s-1|n_f)=\frac{\beta n_{s}}{n_f-n_{s}+1}\mathbb{P}(n_s|n_f),\nonumber\eeq we then obtain:
    \beq\mathbb{P}(n_s |n_f)&\approx&1-\frac{\beta n_{s}}{n_f-n_{s}+1}\\ \mathbb{P}(n_s-1|n_f)&\approx&\frac{\beta n_{s}}{n_f-n_{s}+1}. \label{firstorder1}\eeq
    We ignore all other probabilities in the first order for $\beta\ll 1$. When $n_f\geq n_{s}$ and $\beta \ll 1$ almost all sites are occupied.
\item When there are less TFs than the number of available sites ($0<n_f<n_{s}$), then for $\beta\ll 1$ only $\mathbb{P}(k=n_f-1|n_f)$ and $\mathbb{P}(k=n_f |n_f)$ have a contribution in the leading order of equation (\ref{eigenvector2}). We obtain:
    \beq \mathbb{P}(n_f |n_f)&=&1-\frac{\beta n_f}{n_{s}-n_f+1}\\ \mathbb{P}(n_f-1|n_f)&=&\frac{\beta n_f}{n_{s}-n_f+1}.\label{firstorder2}\eeq
    We neglect all other probabilities in the first order for $\beta\ll 1$.
\end{enumerate}
Combining equations (\ref{probantf}), (\ref{decompnsites}) and the first order approximations in $\beta$, the probability $\mathbb{P}_{n_{s}}$ that all sites are simultaneously occupied is:
\beq
\mathbb{P}_{n_{s}}&=&e^{-\alpha}\sum\limits_{n_f=n_{s}}\limits^{\infty}\left(1-\frac{\beta n_{s}}{n_f-n_{s}+1}\right)
\frac{ \alpha^{n_f}}{n_f!}\label{pnsdepart}.
\eeq
Using the partial sum:
\beq
S(x)=\sum\limits_{k=0}\limits^{n_{s}-1}\frac{x^{k}}{k!},
\eeq
and after some computation (see appendix) we can write:
\beq
\mathbb{P}_{n_{s}}(\alpha)=1-e^{-\alpha}S(\alpha)-\beta n_{s}e^{-\alpha} \int\limits_{0}\limits^{1}
\frac{e^{\alpha u}-S(\alpha u)}{u^{n_{s}}} \,du.\label{alloccupied}
\eeq
For $\beta\ll 1$, $\mathbb{P}_{n_{s}}$ is an increasing function of $\alpha$ and a decreasing function of $\beta$ (see appendix). Increasing the number $\alpha$ of TFs leads to an increase in the
probability that all sites are occupied, while increasing $\beta$ decreases the probability that all sites are occupied.

\begin{figure}
  \subfigure[Probability $\mathbb{P}_k$ for $n_s=2$ and $\beta=\frac{1}{10}$]{\includegraphics[width=6cm]{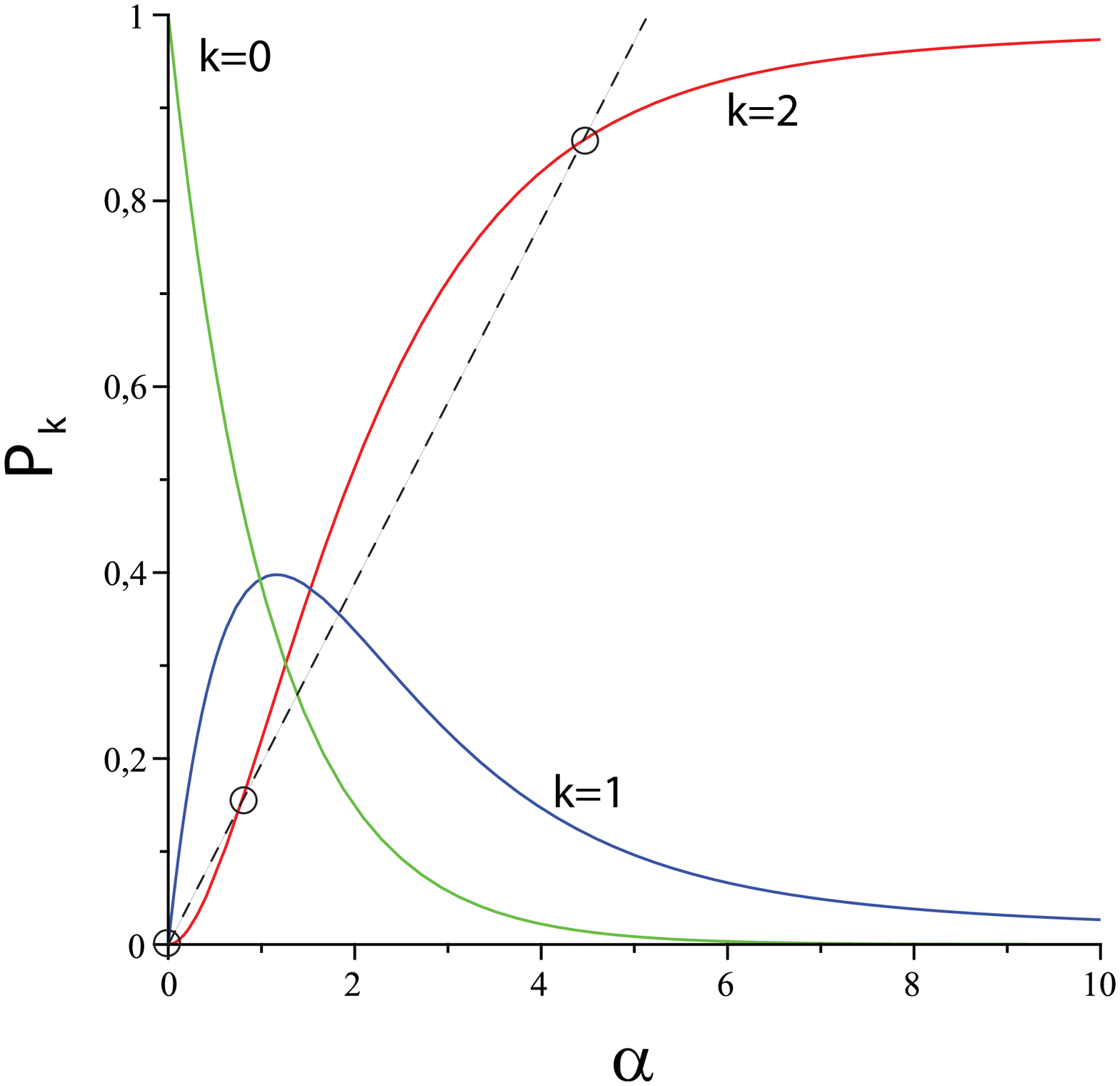}}
  \subfigure[Probability $\mathbb{P}_k$ for $n_s=2$ and $\beta=1$ ]{
\includegraphics[width=6cm]{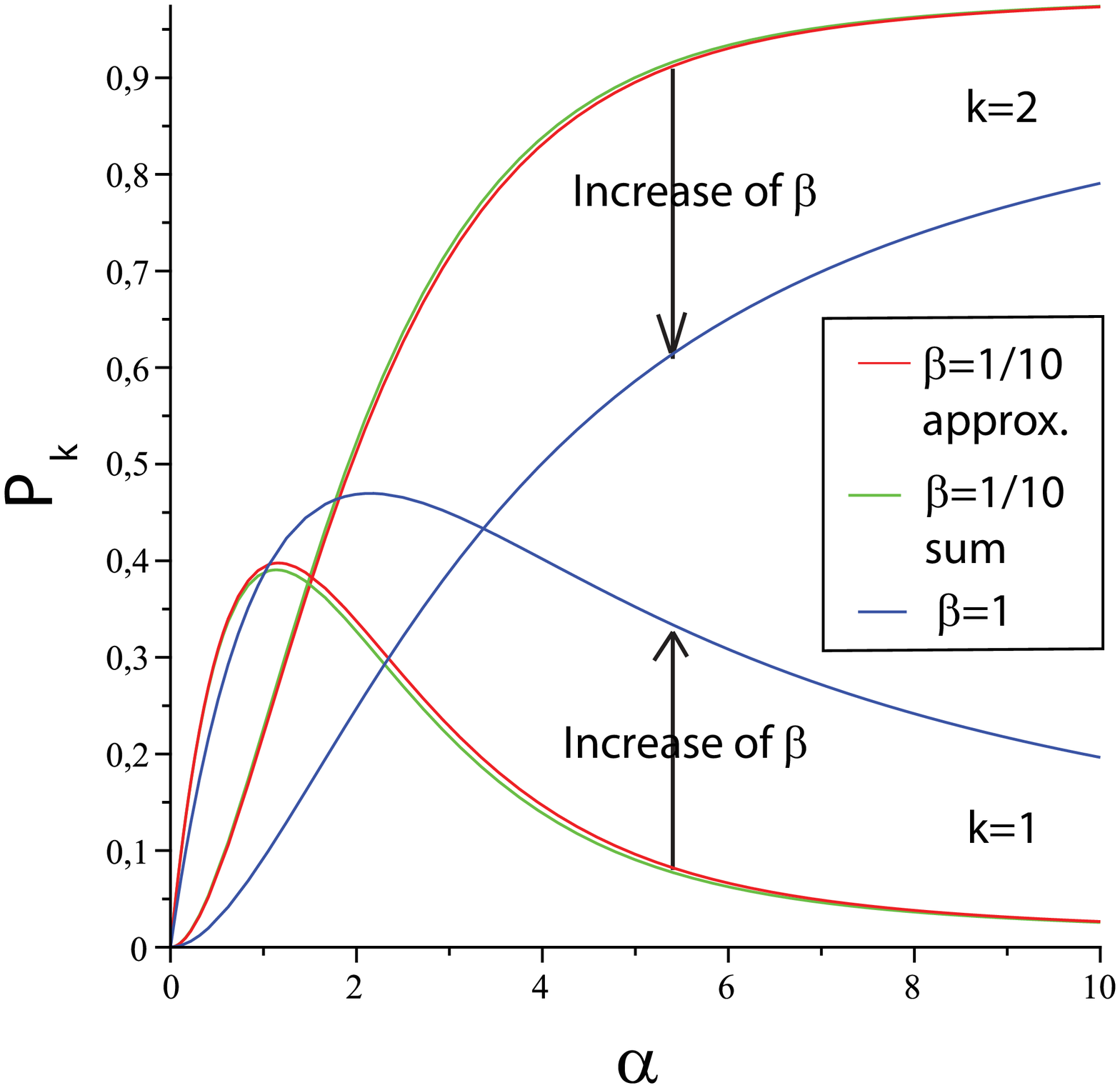}}
  \caption[(a) \textit{$\mathbb{P}_k$  as a function of $\alpha$ for $n_s=2$}\newline(b) \textit{Impact of a change in $\beta$}]{ \textbf{$\mathbb{P}_k$ for $n_s=2$.(a) $\mathbb{P}_{k}$ as a function of $\alpha$ for $\beta=\frac{1}{10}$}. $\mathbb{P}_k$ is computed through approximation (\ref{probamultiplefinal}). For a TF activating its own transcription when both sites are simultaneously occupied, the two stable values for $\alpha$ (high and low values for $\alpha$) and the unstable value (in the middle) are represented along the dotted line. \textbf{(b) The impact of a change in $\beta$.} Curves in red and green are for $\beta=1/10$, the curve in blue is for $\beta=1$.
For $\frac{k_{a}}{k_{d}}=20\, \mu$Mol$^{-1}$
\cite{Baliga} and $\beta=\frac{1}{10}$ when $C=0$, this corresponds
to a ligand concentration of $0\, \mu$Mol (red and green) and $0.5\, \mu$Mol in blue. The curve in red is computed through  through approximation (\ref{probamultiplefinal}). Curves in blue and green are computed through finite sums of (\ref{pkcomplete})
(200 terms).
}\label{2possiblesites}
\end{figure}

\begin{figure}
\subfigure[Probability $\mathbb{P}_k$ for $n_s=4$ and $\beta=\frac{1}{10}$ ]{\includegraphics[width=6cm]{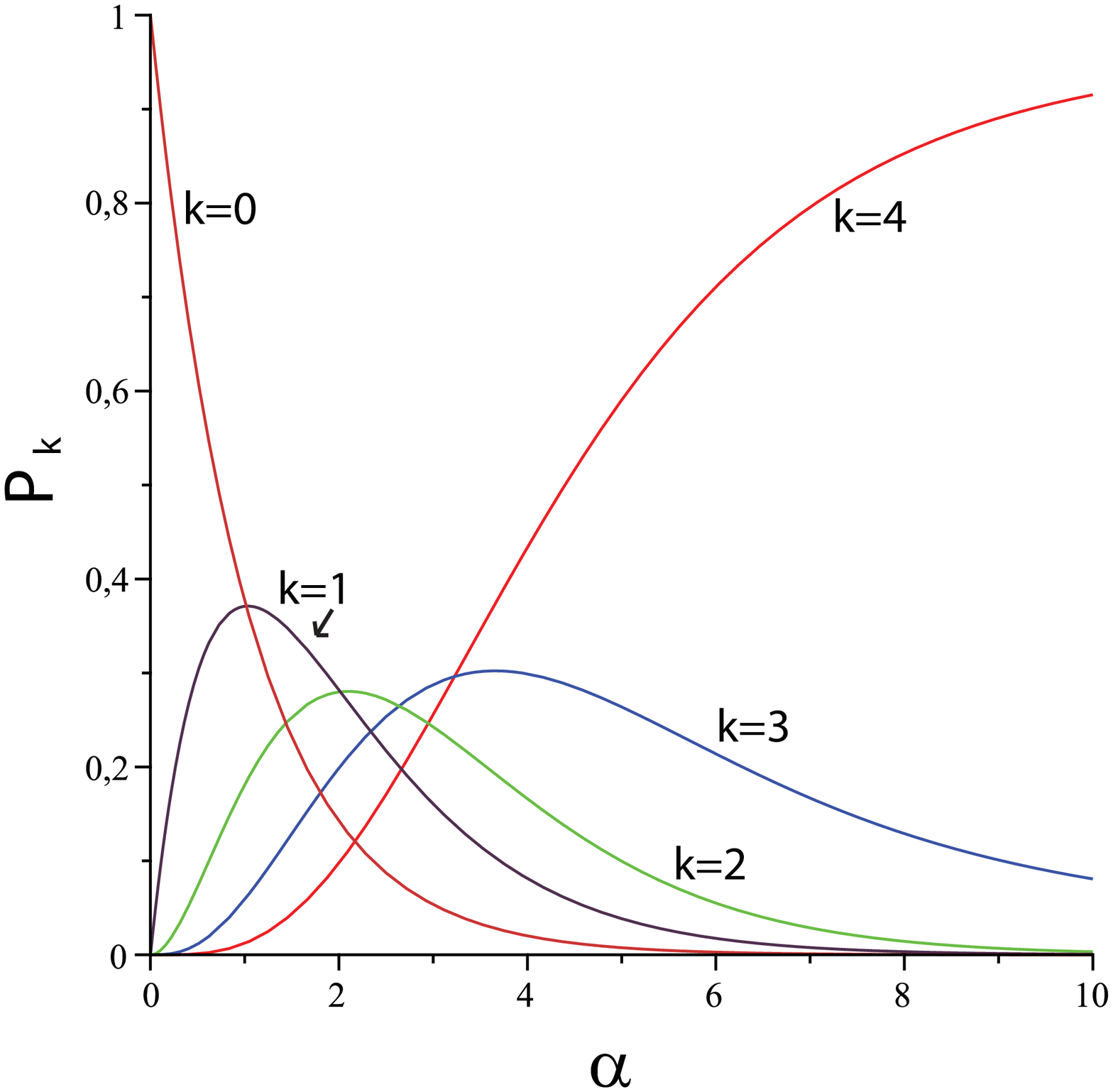}}
\subfigure[Probability $\mathbb{P}_k$ for $n_s=4$ and $\beta=1$ ]{
\includegraphics[width=6cm]{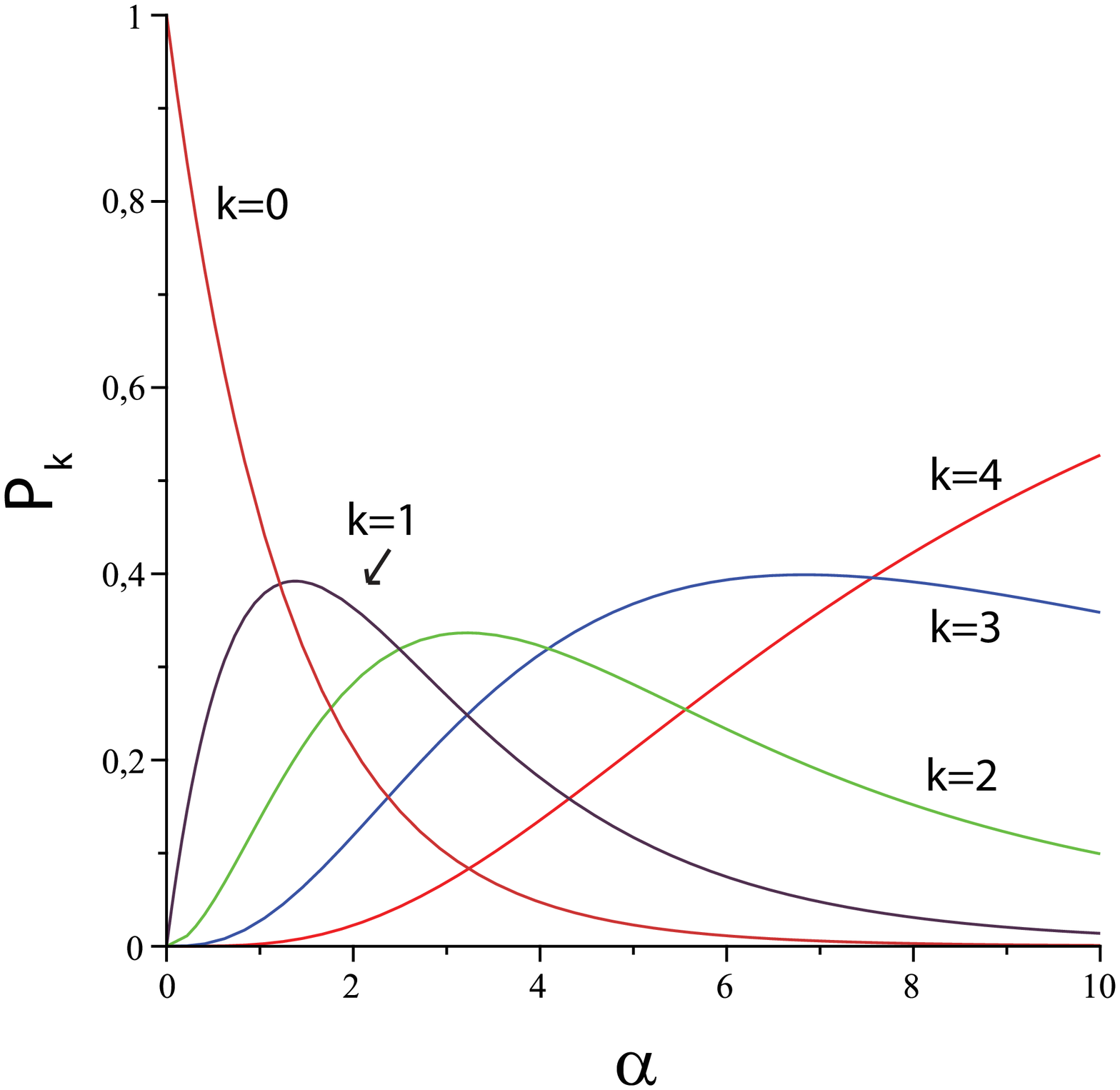}}
  \caption[(a) \textit{$\mathbb{P}_k$  as a function of $\alpha$ for $n_s=4$}\newline(b) \textit{Impact of a change in $\beta$}]{\textbf{ $\mathbb{P}_k$ for $n_s=4$ as a
function of $\alpha$ for
(a) $\beta=1/10$  and (b) $\beta=1$}. $\mathbb{P}_k$ is computed through finite
sums of (\ref{pkcomplete}) (200 first terms).  With
$\beta=\frac{1}{10}$ when $C=0$ and $\frac{k_{a}}{k_{d}}=20\,
\mu$Mol$^{-1}$ \cite{Baliga}, this corresponds to a ligand
concentration of $0\, \mu$Mol (left) and $0.5\, \mu$Mol
(right).}\label{compet4}
\end{figure}

Similarly (see appendix), with equations (\ref{middleoccuppied}), (\ref{allbutoneoccupied}) and (\ref{alloccupied}) the asymptotic expression of the occupation ratio $\mathbb{P}_{k}$ for $\beta\ll 1$ is given by:
\beq
\mathbb{P}_{k}
=\left\{
\begin{array}{l l}\ds
1-e^{-\alpha}S(\alpha)-\beta n_{s}e^{-\alpha}
\int\limits_{0}\limits^{1}
\frac{e^{\alpha u}-S(\alpha u)}{u^{n_{s}}} \,du  &\mbox{ for } k=n_s\\\ds
e^{-\alpha}\frac{\alpha^{n_{s}-1}}{(n_{s}-1)!}\left(1-\beta\frac{n_{s}-1}{2}\right)+\beta
n_{s}e^{-\alpha} \int\limits_{0}\limits^{1}
\frac{e^{\alpha u}-S(\alpha u)}{u^{n_{s}}} \,du &\mbox{ for } k=n_s-1\\\ds
e^{-\alpha}\frac{\alpha^{k}}{k!}\left( 1+\beta\left(\frac{\alpha
}{n_{s}-k}-\frac{k}{n_{s}-k+1} \right) \right) &\mbox{ for } k\leq
n_s-2.\end{array}\right.\label{probamultiplefinal}
\eeq
We plot in figure
\ref{2possiblesites} (resp.
\ref{compet4}) the function $\mathbb{P}_{k}$ as a function of $\alpha$ for $n_s=2$
(resp. $n_s=4$).

\subsection{Consequence of the analysis for gene expression stability} \label{section3}
We now use our results on the occupation ratios $\mathbb{P}_{k}$ to
show that at least two binding sites are required for the regulation
of a genetic switch. A genetic switch is a special type of
autoregulated gene. These autoregulated genes code for TFs that
regulate the transcription of their own gene. A genetic switch is a
process allowing two stable values for the transcription
rate of a gene \cite{Ptaschnebook}: an "on" position where
the gene is transcribed and an "off" position where it is not
transcribed. If the gene is "on", transcription is maintained at
high levels through autoregulation. If the gene is "off",
transcription remains at low levels and does not turn on without an
external signal. Genetic switches play a central role in cellular
differentiation, memory and plasticity \cite{Wang,Crews}.

We shall show that a bistable genetic switch can only appear when there are at least two binding sites regulating the transcription of the gene. We first determine the steady state concentration of the TF due the balance of production and degradation by enzymes. For a gene transcribed at a rate $r$ when there are $N_{on}$ occupied binding sites, the steady state production $\lambda$ of TFs verifies:
\beq
\lambda=r \mathbb{P}_{N_{on}}=rf(\alpha),
\eeq
where $f(\alpha)=\mathbb{P}_{N_{on}}(\alpha)$ is given by formula
(\ref{probamultiplefinal}) and depends on $N_{on}$ and $n_s$. The
steady state value $\alpha=\frac{\lambda}{K}$ satisfies the
nonlinear equation:
\beq  \label{bistabprod}
Rf(\alpha)=\alpha,
\eeq
where $R=\frac{r}{K}$. Bistability appears when equation
(\ref{bistabprod}) has two stable solutions, thus equation
(\ref{bistabprod}) must have three solutions (two stable and one
unstable in between). The number of solutions depends on the
parameters $N_{on}$ and $R$: For $N_{on}=0$, as plotted in figures
\ref{2possiblesites} and \ref{compet4}, $f(\alpha)=\alpha/R$ has
only one solution. For $N_{on}=1$, $f(\alpha)=\alpha/R$ has one
solution for $R$ small and two solutions for $R$ large. For
$N_{on}\geq 2$, using formula (\ref{probamultiplefinal}) and as
plotted in figures \ref{2possiblesites} and \ref{compet4}, $f$ is a
sigmoid type function. For $R$ sufficiently large, equation
(\ref{bistabprod}) has three solutions (figure \ref{2possiblesites})
and two of them are stable. A gene following such activation
properties is a bistable switch. Conversely, for $R$ sufficiently
small, $\alpha \approx 0$ is the only stable solution. The critical
value of $R$ can be characterized geometrically, as the point where
$\alpha/R$ is tangent to $f(\alpha)$. For this critical value there
is a stable point at the origin and a saddle point at the tangent
point. To conclude, a bistable switch requires at least two binding
sites regulating the gene and the parameter $R$ must be sufficiently
large.

\section{Formation of the Hunchback boundary by the Bicoid gradient}
\begin{figure}
\centerline{
\subfigure[$\alpha_h(x)$ for different values of B]{\includegraphics[width=6cm]{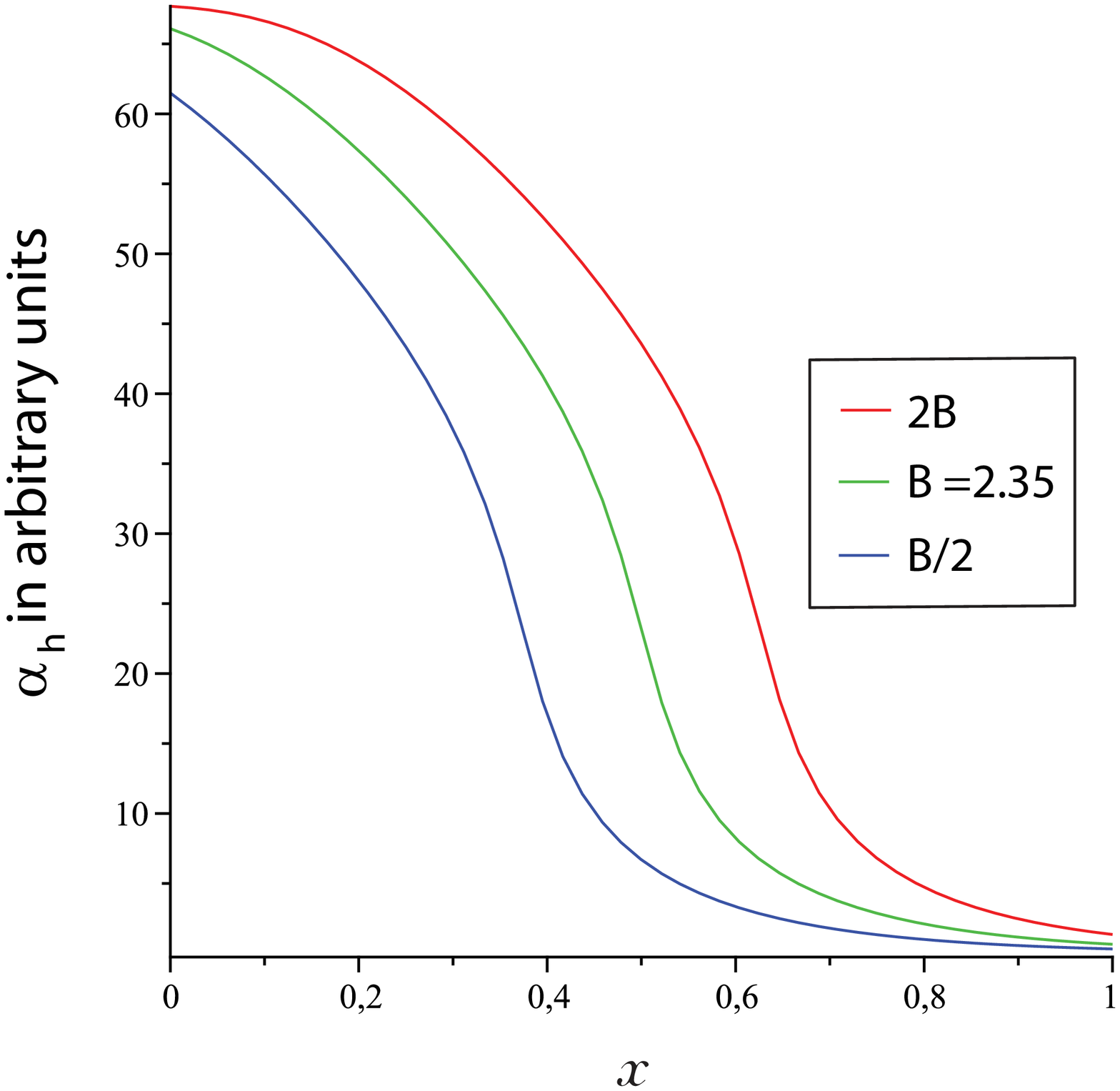}}
\subfigure[$\alpha_h(x)$ for different values of R]{\includegraphics[width=6cm]{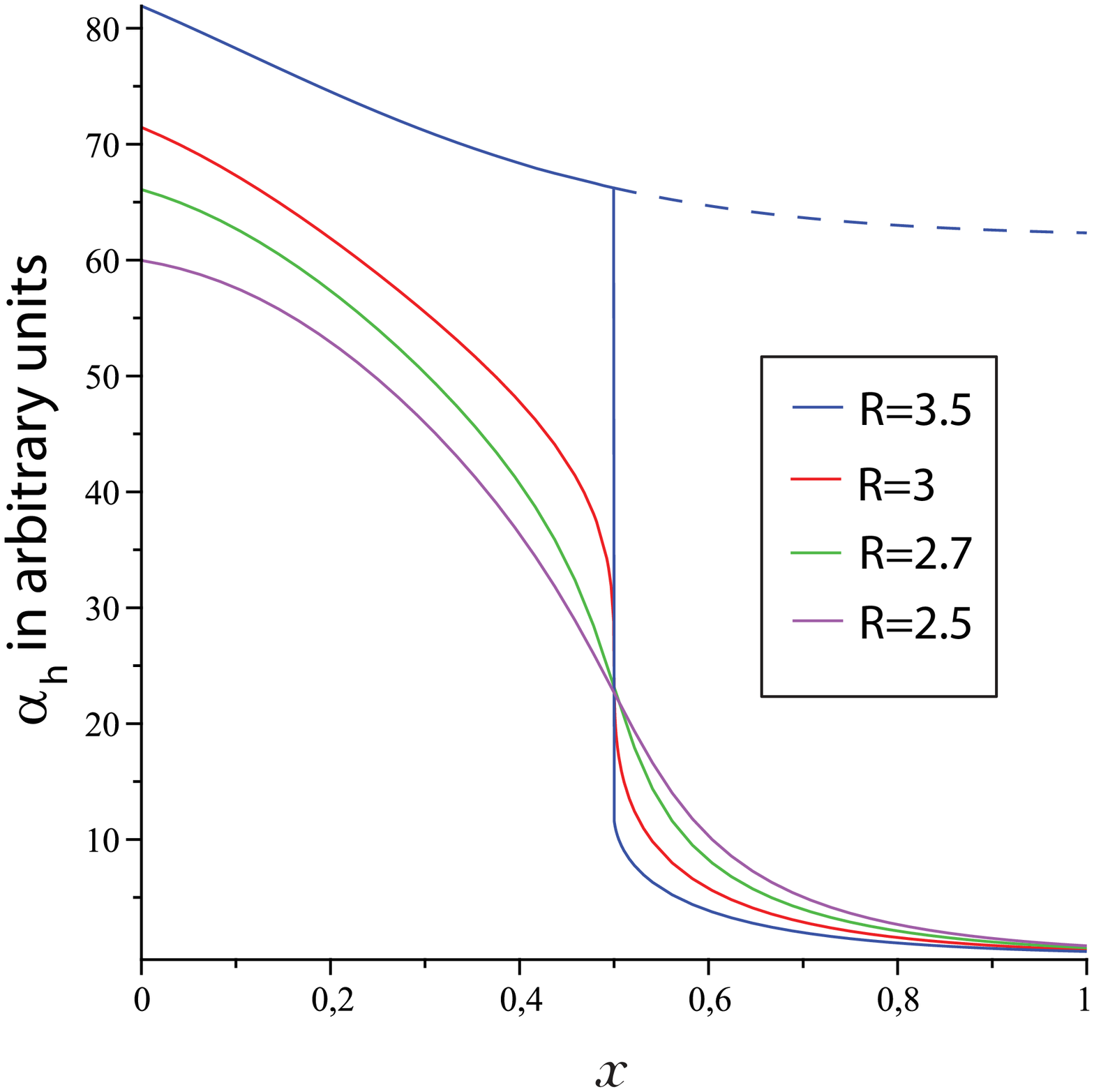}}}
\centerline{
\subfigure[$R(1-\mathbb{P}_{0,b}(1-\mathbb{P}_{2,h}))-\alpha_h$ as a function of $\alpha_h$]{\includegraphics[width=6cm]{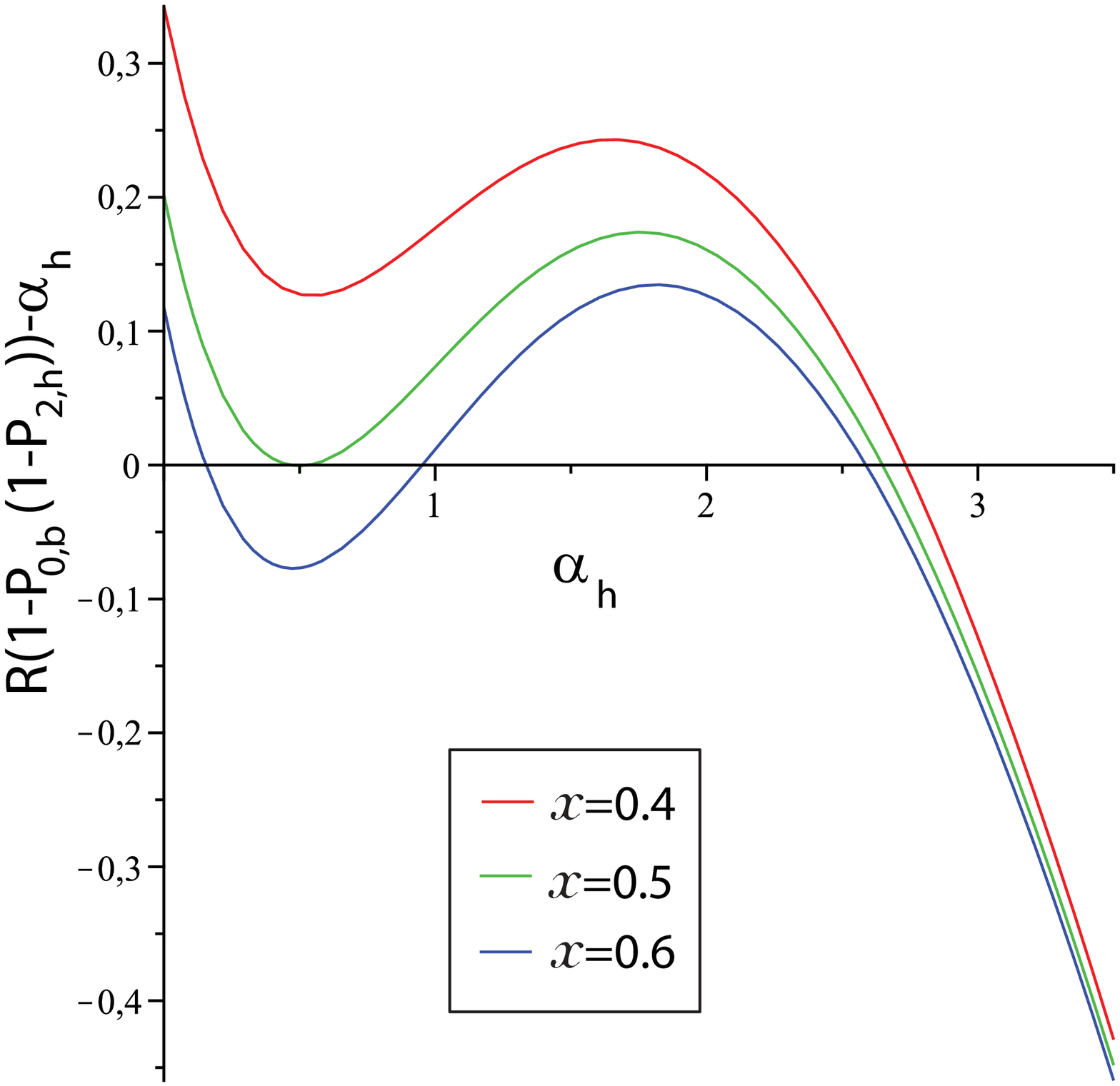}} \subfigure[Bifurcation diagram of $\alpha_h(x)$]{\includegraphics[width=6cm]{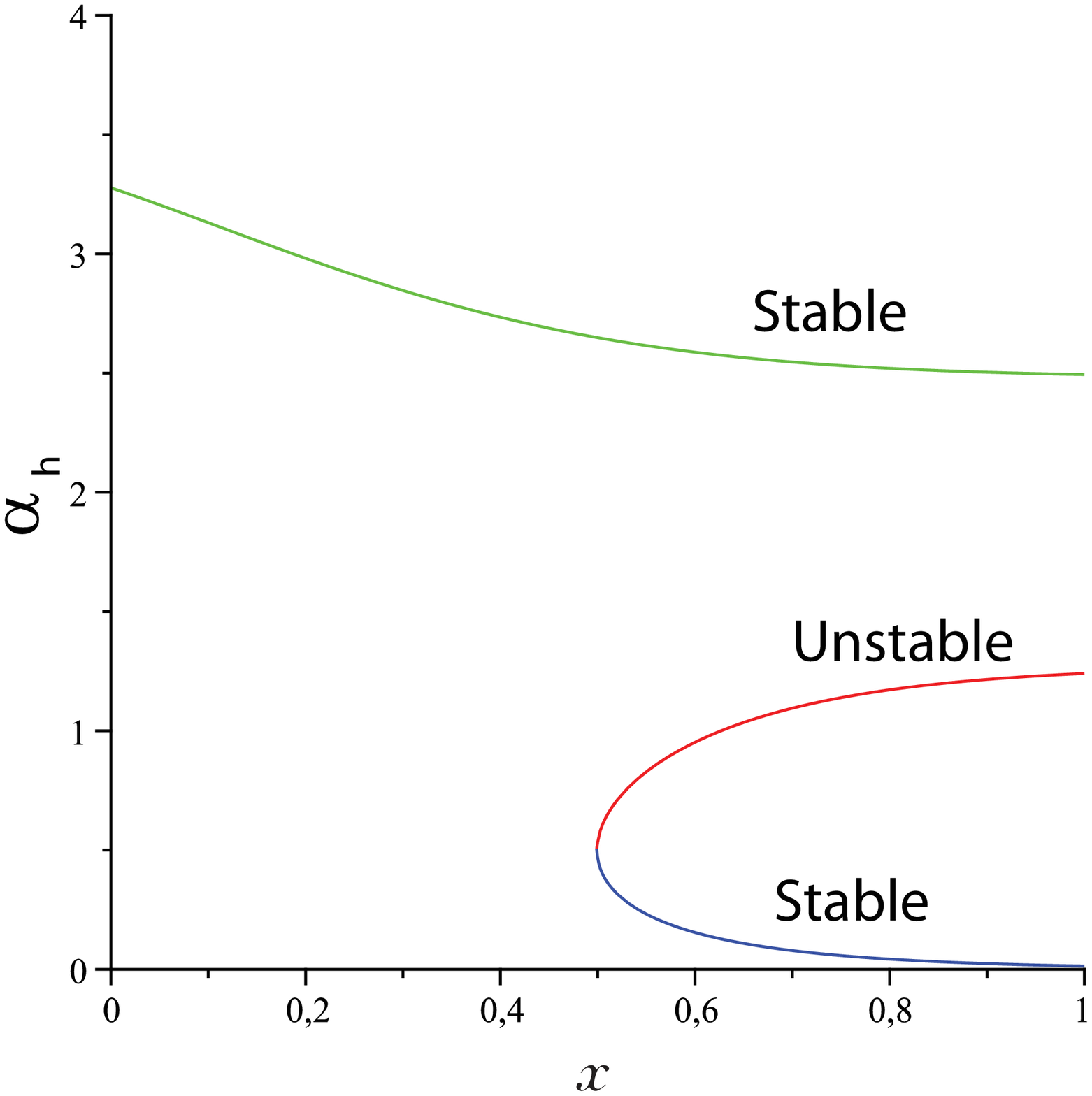}}} \caption[(a) \textit{Hb concentration for different values of $B$}\newline(b) \textit{Hb concentration for different values of $R$}\newline(c) \textit{$R(1-\mathbb{P}_{0,b}(1-\mathbb{P}_{2,h}))-\alpha_h$ as a function of $\alpha_h$} \newline (d) \textit{Bifurcation diagram of $\alpha_h(x)$ for $R=3.5$}]{\textbf{(a) Hb concentration $\alpha_h(x)$ for different values of $B$:} $B=2.35/2$ (blue), $B=2.35$ (green) and $B=2.35*2$ (red). Here, we use $R=2.7$. All curves for $\alpha_h$ where scaled by a factor 25 to
obtain the same numerical values as the the concentration in arbitrary units for in vivo data reproduced in figure \ref{hbbcd}a. \textbf{(b) $\alpha_h(x)$ for different values of $R$:} $R=2.5$ (monostable), $R=2.7$ (monostable, value from \cite{Lopes}), $R=3$ (critical value for bistability) and $R=3.5$
(bistable). For $R=3.5$, there are two stable points: the high (dotted lines) and the low (continue line) stable value. $B$ in (\ref{expobcd}) was adjusted for each of the curves to cut 25 in $x=0.5$: $B=3.3$ for $R=2.5$, $B=2.35$ for $R=2.7$, $B=1.58$ for $R=3$ and $B=0.95$ for $R=3.5$. \textbf{(c) $R(1-\mathbb{P}_{0,b}(\alpha_b(x))(1-\mathbb{P}_{2,h}(\alpha_h)))-\alpha_h$ as a function of $\alpha_h$ for $R=3.5$.} The curves are for $x=0.4$, 0.5 and 0.6. We use $B=0.95$ as in figure \ref{hbsanshkb}b. \textbf{(d) Bifurcation diagram of $\alpha_h(x)$.} This bifurcation diagram is given by the solutions of $R(1-\mathbb{P}_{0,b}(\alpha_b(x))(1-\mathbb{P}_{2,h}(\alpha_h)))-\alpha_h=0$ as a function of $x$. We use  $B=0.95$ and $R=3.5$ as in figure \ref{hbsanshkb}c.}
\label{hbsanshkb}
\end{figure}

We shall now apply our analysis to determine the formation of the
Hunchback TF (hb) boundary by the Bicoid (bcd) morphogen gradient in
the drosophila embryo. The bcd gradient regulates a number of
downstream TFs involved in the gap gene network
\cite{Feng,Ashyraliyev}, which determine the position of body sections along the
anterior-posterior (A-P) axis in the drosophila embryo. Among these
gap genes, hb is responsible for thoracic development
\cite{Feng,Ashyraliyev}. Given a bcd gradient, we propose to
determine the spatial distribution of hb. Our analysis shows how a
broad bcd gradient can trigger a sharp transition in the hb density
in the middle of the embryo. We reproduce the bcd and hb density
measured in vivo \cite{Feng} in figure \ref{hbbcd}a. To distinguish
the values of $\alpha$ and $\beta$ for the hb and bcd TFs required
in our previous model, we shall use subscript $_h$ for the hb TF and
$_b$ for the bcd TF. We approximate bcd gradient as exponential
\cite{Feng}:
\beq  \label{expobcd}
\alpha_b(x)=Be^{-kx},
\eeq
where $x\in[0,1]$ is the normalized A-P position ($x=X/L$ where L is
the length of the drosophila embryo). We use $k=5.5$, corresponding
to the best fit for the in vivo data \cite{Feng}. The constant $B$
cannot be obtained directly from in vivo data. However, since
\beq
\alpha_b=e^{-k\left(x-\frac{\ln(B)}{k}\right)},
\eeq
changing the value of $B$ is equivalent to an x-translation of the
hb and bcd densities. We choose $B$ such as the hb boundary is in
the middle of the drosophila embryo (see figure \ref{hbsanshkb}b).

hb transcription results from the binding of the hb TF and the bcd
TFs to a promoter with 6 bcd binding sites and 2 hb sites
\cite{Lopes,Struhl,Driever}. Hb is transcribed at a
rate $r$ when there are two hb or at least one bcd bound to the
sites, described as
\beqq \label{production1}
\emptyset\xrightarrow[\mbox{\scriptsize  rate }r]{\mbox{\scriptsize At least 1 bcd bound}}hb\\
\emptyset\xrightarrow[\mbox{\scriptsize  rate }r]{2 \mbox{\scriptsize  hb bound}}hb .\\
\eeqq
The hb density is proportional to
the steady state production of hb given by $\lambda=r(1-P)$, where
\beq
P=\mathbb{P}_{0,b}(1-\mathbb{P}_{2,h})
\eeq
is the probability that hb is not transcribed, $\mathbb{P}_{0,b}$ is
the probability that no bcd are bound to the promoter and
$1-\mathbb{P}_{2,h}$  the probability that there are not two hb
bound. At equilibrium, using $\alpha_h=\frac{\lambda}{K}$, we obtain
the steady state equation
\beq
\alpha_h=\frac{\lambda}{K}= R
(1-\mathbb{P}_{0,b}(1-\mathbb{P}_{2,h})),
\label{Pequaprobas}
\eeq
where $R= \frac{r}{K}$ and $K$ is the degradation constant for hb.
Equation (\ref{Pequaprobas}) is implicit for the mean number
$\alpha_h$ of hb, that we shall now compute. We will now evaluate
separately expressions $\mathbb{P}_{0,b}$ and $\mathbb{P}_{2,h}.$
Along the A-P axis parameterized by the position $x$,
$\mathbb{P}_{0,b}$ depends on the mean number $\alpha_b(x)$ of bcd
TFs and on the ratio $\beta_b=\frac{\overline{T}_S}{\overline{T}_b}$
of the search time of bcd over the binding time. {To evaluate
$\beta_b$, we use the binding reaction of a bcd to its target site
$S$:}
\beq
S+bcd\rightleftharpoons S.bcd\label{sb},
\eeq
where $S.bcd$ is the bcd TF bound to its target site.  The
equilibrium constant $K_d=\frac{[S.bcd]}{[S][bcd]}$ is the ratio of
the forward to the backward rate of
(\ref{sb}), equivalently:
\beq
K_d=\frac{\overline{T}_S}{\overline{T}_bN_aV}=\frac{\beta_b}{N_aV}.
\eeq
For $K_d=0.24 nM$ \cite{Burz}, a nucleus of volume $V\approx1\mu
m^3$ and with $N_a$ the Avogadro number we obtain $\beta_b=K_d N_a
V.\approx0.14. $

\begin{figure}
\subfigure[Boundary in the density of the autoregulated TF.]{\includegraphics[width=6cm]{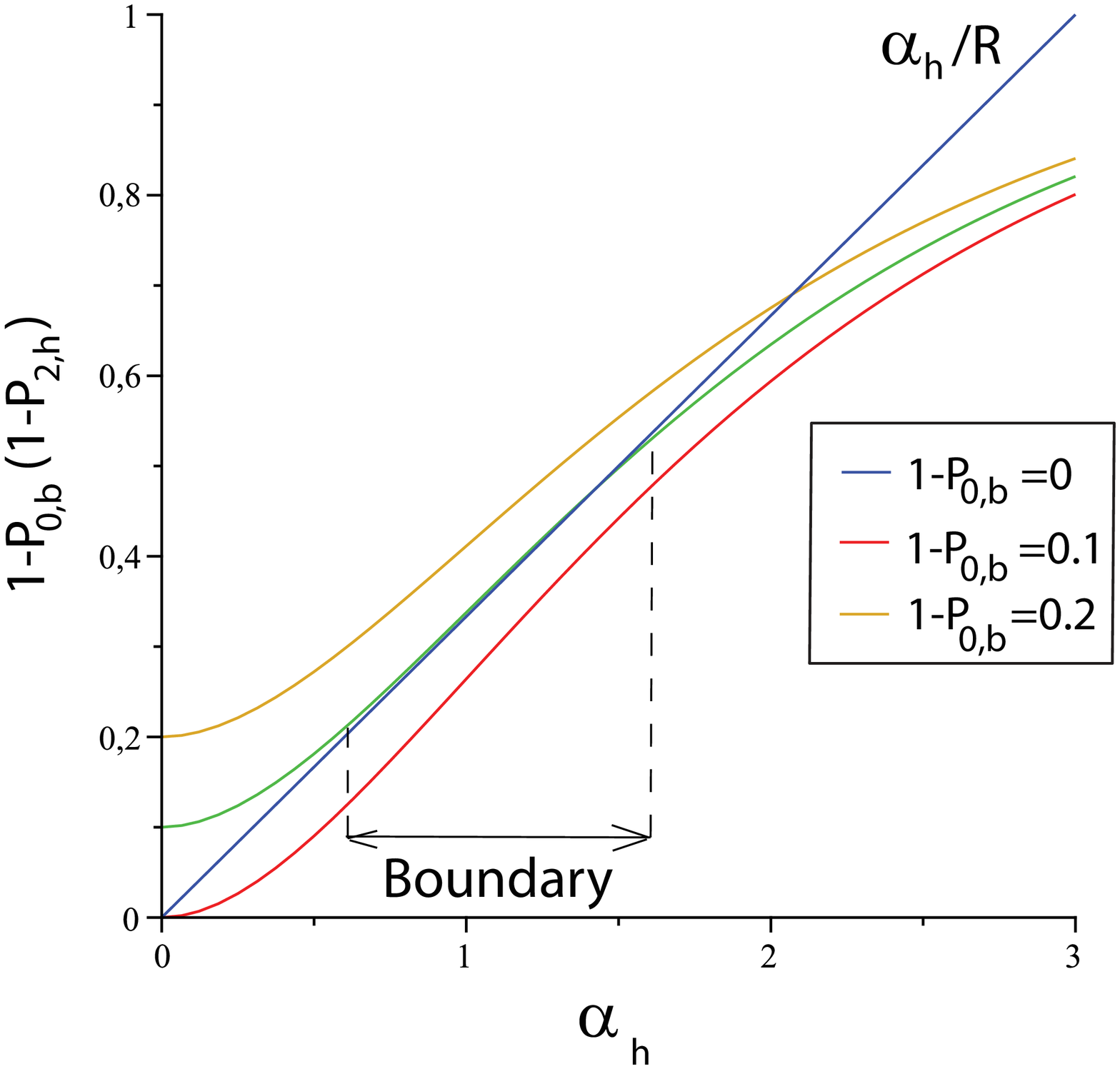}}
\caption[\textit{Boundary in the density of the autoregulated TF.}]{\textbf{Boundary in the density of the autoregulated TF.} Are representd $\alpha_h/R$ (blue)
and the proportion of time $1-\mathbb{P}_{0,b}(1-\mathbb{P}_{2,h}(\alpha_h))$ the hb gene is active for $P(x)=1-\mathbb{P}_{0,b}=0$ (blue), $0.1$ (red) and $0.2$ (yellow). The curves are all for the critical value $R=3$ to amplify the boundary in hb. We use $B=1.58$ as in figure \ref{hbsanshkb}b. The boundary comes from $1-\mathbb{P}_{0,b}(1-\mathbb{P}_{2,h}(\alpha_h))$ which is tangent to $\alpha_h/R$ at the point where $\mathbb{P}_{2,h}(\alpha_h)$ changes concavity. A small variation in $P(x)$ then induces a
large variation in $\alpha_h$.
}
\label{boundary}
\end{figure}

To compute $\mathbb{P}_{0,b}$, we use formula
(\ref{probamultiplefinal}) with $k=0, n_s=6$ and obtain:
\beq
\mathbb{P}_{0,b}=e^{-\alpha_b}\left(1+\frac{\beta_b
\alpha_b}{6}\right).
\eeq
We shall now evaluate the probability $1-\mathbb{P}_{2,h}$. In the absence of any precise data on the dissociation constant of hb from its binding site, we consider that binding is fast enough so that $\beta_h\approx0$. Using expression (\ref{probamultiplefinal}) for the probability $\mathbb{P}_{2,h}$ with $k=n_s=2$, we obtain
\beq
1-\mathbb{P}_{2,h}=e^{-\alpha_h}+\alpha_he^{-\alpha_h}\label{equa123456}.
\eeq
Finally, at steady state, the equilibrium condition
(\ref{Pequaprobas}) reads:
\beq \label{equa12345}
R \left(1-e^{-\alpha_b}\left(1+\frac{\beta_b \alpha_b}{6}\right)\left(e^{-\alpha_h}+\alpha_he^{-\alpha_h}\right)\right)=\alpha_h.
\eeq
We solve equation (\ref{equa12345}) numerically (with Maple) to
express $\alpha_h$ as a function of $\alpha_b$. We plot in figure
\ref{hbsanshkb}a-b several solutions associated with different values of $R$ and
$B$. As pointed out in equation (\ref{expobcd}) and plotted in
figure \ref{hbsanshkb}a, changing the value of $B$  is equivalent to
a x-translation of the hb and bcd densities. To further study the
different types of solutions, we will vary the parameter $R$.
Following the discussion in section \ref{section3} on bistability,
for $N_{on}=n_s=2$  the dynamics for hb can potentially be bistable.
We show now that for $n_s=2$ and $R<3$, hb is always monostable. To
compute the critical value $R_c$ after which bistability occurs, we
shall use the functions:
\beq
P(x)&=&1-\mathbb{P}_{0,b} \\
f(\alpha_h)&=&\mathbb{P}_{2,h},
\eeq
where $P(x)$ depends on  $x$ through  $\alpha_b$ (\ref{expobcd}).
The function $f(\alpha_h)$ is the fraction of time hb is
autoactivated by the hb and $P(x)$ is the fraction of time the gene
is activated by the bcd gradient. Equation (\ref{equa12345}) can then
be rewritten as:
\beq
R(1-(1-P(x))(1-f(\alpha_h)))=\alpha_h\label{erty}.
\eeq
We determine in the appendix the critical value for bistability
given by $R_c=3$. For $R<R_c$ the gene is always monostable, while
for $R>R_c$ the gene is bistable for some values of $P(x)$ and
monostable for others:
\begin{itemize}
\item For $R>3$, hb is monostable for $x<x_c$
and bistable for $x>x_c$ where $x_c$ is a critical position. We
represent the bifurcation diagram in figure \ref{hbsanshkb}d.
Changing $B$ is equivalent to an x-translation in the hb profile and
thus $B$ can be adjusted such as the bifurcation point is $x_c=0.5$
for example. If at time $t=0$ there is no hb, the hb density
converges to the lower stable value, as represented in figure
\ref{hbsanshkb}b. Nevertheless, for a bistable hb dynamic, cells located in $x>x_c$ can switch from the low to the high stable value for a sufficient  perturbation. In the absence of a repressor of hb on the posterior side of the embryo, these cells would stay in the high stable state.
\item For $R<3$, hb is always monostable. When $R$ becomes close to 3,
there is already a boundary in the hb density (figure
\ref{hbsanshkb}b).  This boundary can be characterized by the point
where $f(\alpha_h)$ changes concavity and becomes tangent to a
linear function (figure \ref{boundary}). At the point of concavity
change, a small variation in $P(x)$ induces a large variation in
$\alpha_h$ which produces a sharp transition in the hb density.
\end{itemize}
We conclude that for an auto-regulated hb gene, when the bifurcation
parameter $R$ is close but smaller than the critical value $R_c$,
there is a sharp boundary of hb in the embryo and this boundary does
not require a repressor in the posterior half of the embryo. As
shown in figure \ref{boundary}, at the boundary hb synthesis is
essentially due to autoactivation of hb (the activation $P(x)$ due
to bcd is $\approx 10\%$ whereas the gene is autoactivated
$\approx40\%$ of the time). To obtain a numerical estimation of $R$,
we use the synthesis rate $r$ generated by two hb bound to the
target sites and the degradation rate $K$ of hb. Using the values
from the supplementary material of \cite{Lopes}, $r\approx 19$ and
$K\approx 7.08$ and we obtain
\beq
R \approx 2.7.
\eeq
For $R=2.7$, we observe a steep transition of the hb density at the
middle of the embryo as in the in vivo data from
\cite{Feng} reproduced in figure
\ref{hbbcd}a. The main difference between the theoretical density
(figure \ref{hbsanshkb}b) and the in vivo data from \cite{Feng}
(figure \ref{hbbcd}a) is in the anterior edge where our model leads
to an increase of the hb density instead of a decay as observed in
vivo. This decay in the hb density at the anterior edge of the
embryo is due to a repressive effect induced by the huckebein TF
(hkb) \cite{Ashyraliyev} which we did not model in (\ref{equa12345})
and we shall examine now.

\subsection*{Refining the gradient using hkb repressor}
\begin{figure}\centerline{\subfigure[In vivo concentration of bcd and hb]{\includegraphics[width=6cm]{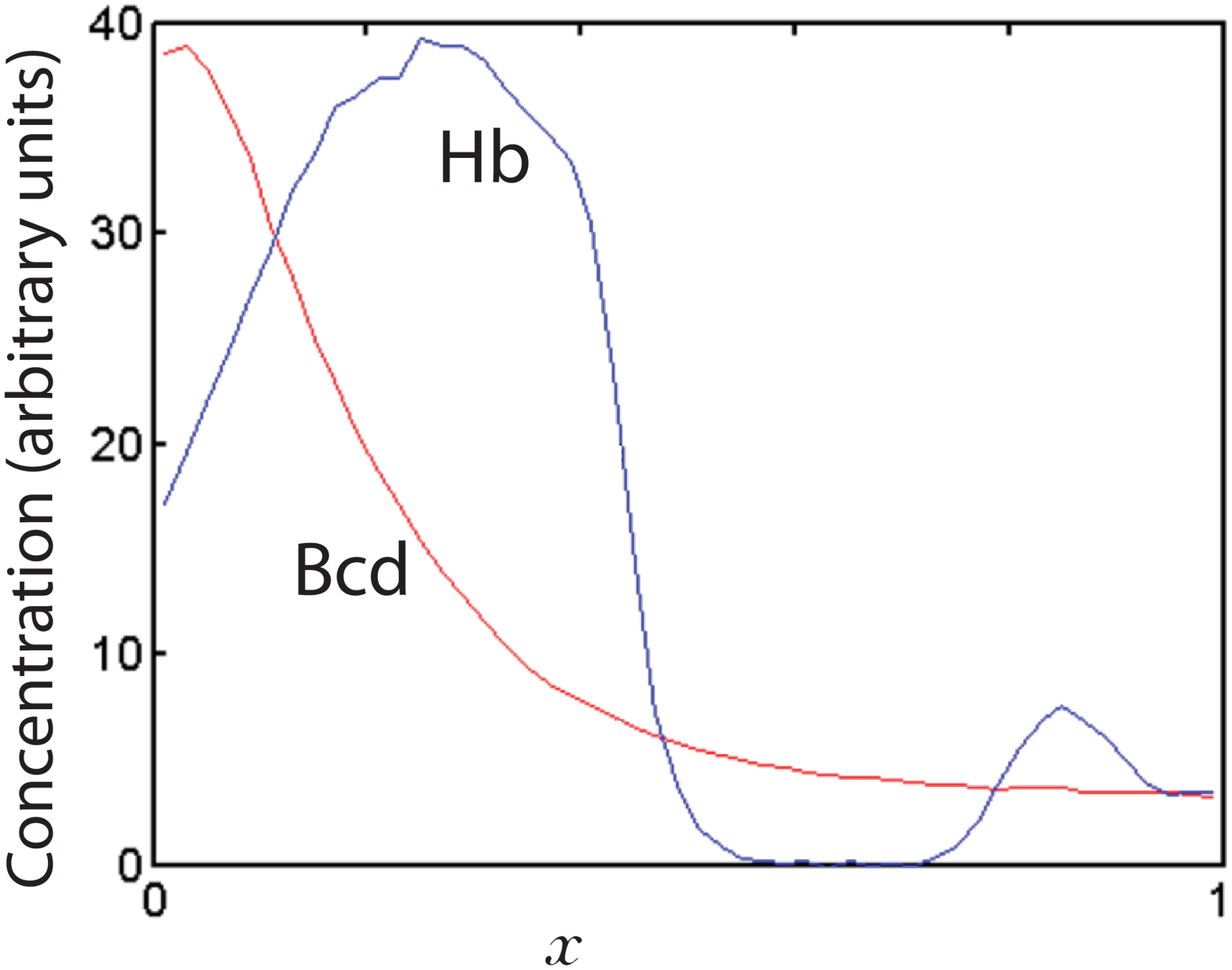}}
\subfigure[Theoretical and experimental concentrations of hb] {\includegraphics[width=6cm]{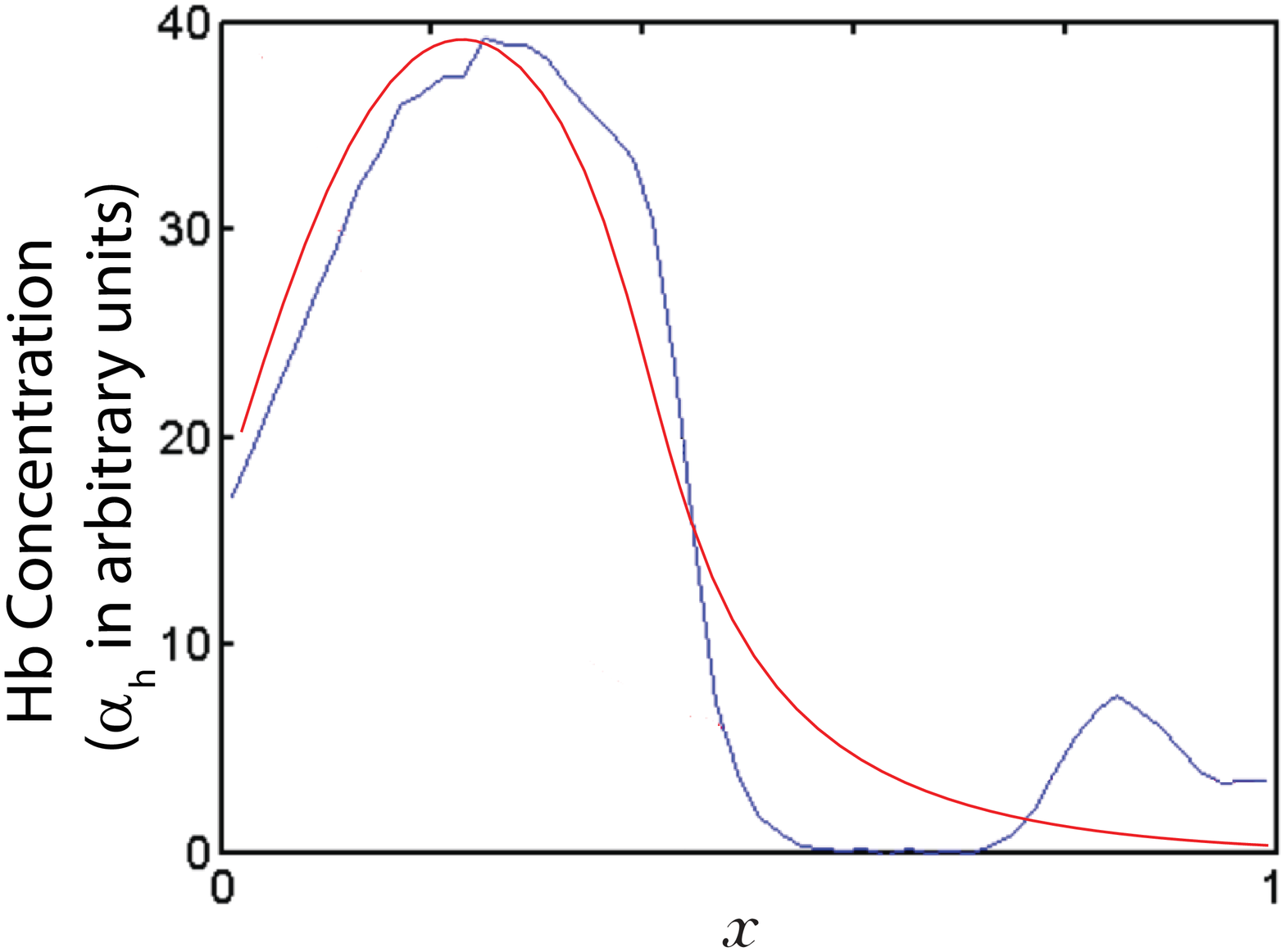}}
}\centerline{ \subfigure[Exponential fit of $\alpha_{hkb}(x)$]{\includegraphics[width=6cm]{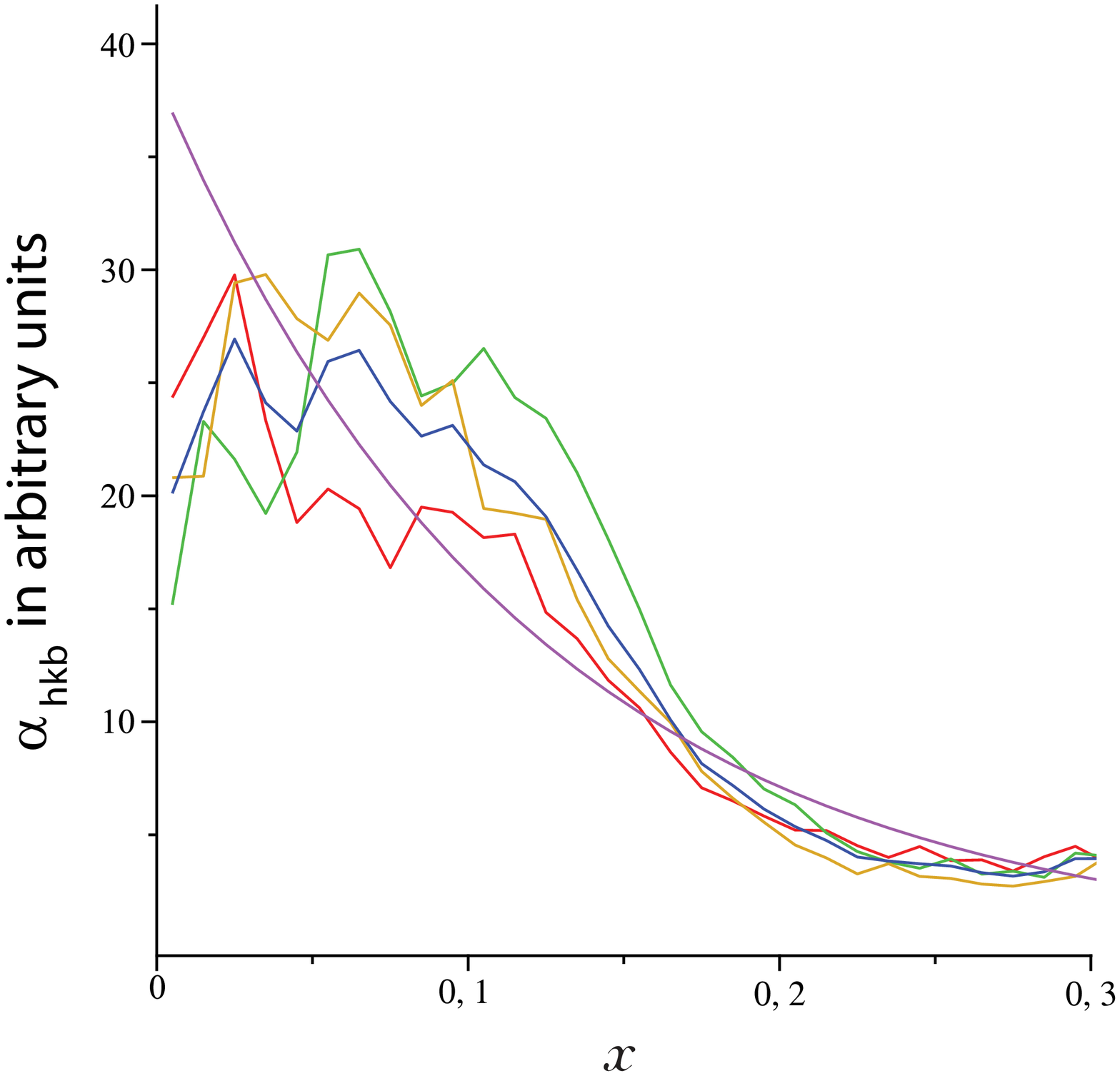}}
\subfigure[$\alpha_h(x)$ for different values of C]{\includegraphics[width=6cm]{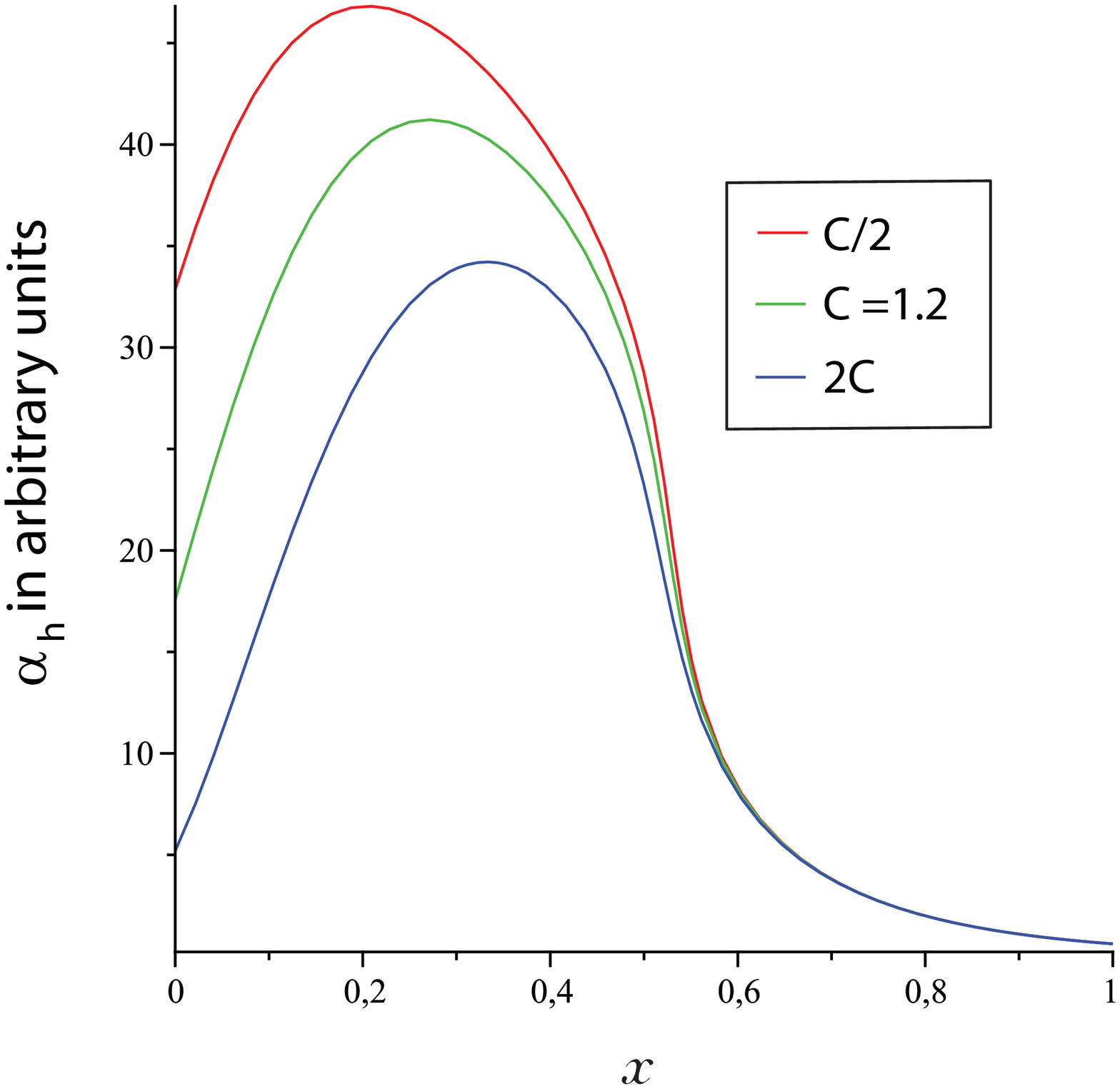}}}
\caption[(a) \textit{In vivo concentration of the bcd and hb TFs from \cite{Feng}} \newline(b) \textit{Theoretical and in vivo concentration of bcd.} \newline (d) \textit{In vivo hkb density from
\cite{Ashyraliyev} and exponential fit used in \ref{expohkb}.}
\newline (d) \textit{Bifurcation diagram of $\alpha_h(x)$ for $R=3.5$}
]{{\bf (a) In vivo concentration of
bcd and hb TFs as a function of $x$.} Figure reproduced from \cite{Feng}.  \textbf{(b) Theoretical and experimental hb concentration as a function of $x$.} Here to compute $\alpha_h(x)$ we take into account bcd activation, hb autoregulation and hkb repression. The parameters $B=1.4$ and $C=1.1$ are
used to fit the in vivo data which is reproduced from \cite{Feng}.
\textbf{(c) In vivo hkb density as a function of $x$ and exponential fit used in equation (\ref{expohkb}).} The in vivo data is from the Flex database \cite{Ashyraliyev,Pisarev,Poustelnikova}. (Since the bcd
and hb densities from \cite{Feng} reproduced in figure (\ref{hbbcd})a are for the
beginning of the 14 A cycle of the development of Drosophilia \cite{Feng}, we use the in vivo hkb densities for the first half of the 14 A
cycle (T1 to T4) from the Flex database to fit (\ref{expohkb}).) \textbf{(d) Hb density $\alpha_h(x)$ for different values of $C$: } for C=0.6 (red), 1.2 (green), 2.4 (blue). The value of $C$ is adjusted to have the same value for the hb density at $x=0$ as for
the in vivo data.}
\label{hbbcd}
\end{figure}

We now account for the repression induced by hkb and consider that
the transcription of the hb gene is repressed when at least one hkb
is bound to the promoter site. Similarly to the analysis that lead
us to equation (\ref{Pequaprobas}), we obtain:
\beq
\alpha_h= R\, \mathbb{P}_{0,hkb}(1-\mathbb{P}_{0,b}(1-\mathbb{P}_{2,h}))\label{hkbequa}.
\eeq
where $\mathbb{P}_{0,hkb}$ is the probability that no hkb are bound. We assume hkb binds to its target fast enough and shall consider that $\beta_{hkb}=0$. Finally, $\mathbb{P}_{0,hkb}$ is then given by:
\beq
\mathbb{P}_{0,hkb}=e^{-\alpha_{hkb}}.
\eeq
To evaluate the distribution $\alpha_{hkb}$ we fit the measured hkb distribution \cite{Ashyraliyev} with an exponential function:
\beq
\alpha_{hkb}=Ce^{k_{hkb}x},\label{expohkb}
\eeq
where $k_{hkb}=11.3$ (see figure \ref{hbbcd}c). The value of $C$ can
not be obtained directly from experimental measurements. Changing
the value of $C$ is equivalent to an x-translation of the repression
due to hkb. We calibrated $C$ to have the same value for the hb
density as in the vivo data (fig \ref{hbbcd}a and d). We solve
equation (\ref{hkbequa}) numerically and obtain an hb density
represented in figure \ref{hbbcd}b. This new theoretical density
obtained is now close to the in vivo data (figure \ref{hbbcd}a), in
particular we recover the  sharp boundary of hb.
The main differences between the theoretical and experimental
densities are located at the posterior side of the hb boundary where
we obtain a higher density than the vivo data and at the
posterior edge where the density is lower. The difference  at the posterior side of the hb boundary might be due to repression of hb by the knirps TF
\cite{Ashyraliyev} which is not modelled here. As for the difference
at the posterior edge, this can be due to activation of hb by the
Caudal TF \cite{Ashyraliyev}. Taking into account these two
regulation pathways should lead to a refined analysis of the hb
density.

\section{Conclusion}
In this paper, we studied transcription activation by TFs starting
from the stochastic nature of the search process for a DNA promoter
site. We then applied our computations to estimate the sharp
boundary induced by a smooth gradient of TFs. In the first part, we
focused on the kinetics of the binding of TFs to their target sites
located on the DNA molecule: when the average number of cycles of
free diffusions and DNA bindings before finding the target sites is
large, the search time $T(n_f,n_s)$ is exponentially distributed and
we estimate the mean (relation (\ref{tempsmultpileunsite1})). Next, we
considered the case of a cell receiving a steady state influx of TFs which can be enzymatically degraded. We modeled
the dynamics of the TFs' binding and unbinding their target sites
and we estimated the fraction of time $\mathbb{P}_k$ that $k$ out of
$n_s$ sites are occupied at steady state. For $n_s=1$ we obtain an
explicit expression in equation (\ref{integ1}). For $n_s\geq 1$, the
general expression of $\mathbb{P}_k$ is given by an
equation (\ref{pkcomplete}) and an asymptotic development for
$\beta\ll 1$ is provided in expression (\ref{probamultiplefinal}). We presented
the different occupation ratios in figure \ref{2possiblesites} and
\ref{compet4} for two and four sites respectively. We consider that
the transcription rate is proportional to the fraction of time a
given number of sites are occupied. For a defined TF concentration
entering the nucleus, our model provides a quantitative input-output
relation in terms of the transcription rate.

When we apply our model to the regulation of hunchback by the bicoid
morphogenic gradient, we focus on the sharp boundary in the hb
density at the middle of the embryo. Several mechanisms accounting
for the formation of sharp boundaries have been proposed: Some
mechanisms \cite{Burz,Lebrecht,Ptashne2,Gregorlimits} result from
cooperative binding while others include a bistable gene
\cite{Lopes} or the antagonistic action of a repressor and activator
gradient \cite{Lohr,Wu,Kasatkin,DHolcmangradient}. Here, we use
neither the repression of hb in the posterior half nor the
cooperative binding of bcd, but we show that, in the absence of
these two mechanisms, a smooth morphogenetic gradient can trigger a
sharp boundary for an autoregulated gene. We also show that
bistability of the autoregulated gene is not a requirement and that
sharp boundaries can be generated by monostable autoregulated genes.
We found the critical value for the transcription rate at which a
bifurcation occurs and gave an estimate in equation
(\ref{Rcritical}). We further show that a bistable gene can produce
a sharp boundary from a smooth gradient. Nevertheless, for a
bistable hb, cells located on the posterior side of the embryo can
switch  from a low stable value to a high one in response to a
sufficiently large perturbation. A repressor on the right hand side
of the boundary would then be required to obtain a reliable boundary
position. Our results show that an autoregulated gene close to
bistability is sufficient to produce a sharp boundary.
%

Here, we focused on a minimal mechanism that allows a morphogenetic
gradient to trigger a sharp boundary in an autoregulated gene. In
order to focus on this minimal system that produces sharp
boundaries, neither hb-repression in the posterior half nor
cooperative binding of bcd are modeled. Both repression \cite{Lohr}
and cooperative binding \cite{Lebrecht} are already known to play a
key role in the formation of the sharp boundary of hunchback and it
would thus be interesting to expand our model to take them into
account. With autoregulation, it would then be interesting to see
how these three mechanisms, which appear to be redundant,  produce
sharp and robust boundaries in the embryo.


\section{Appendix}
\subsection{The pdf of $T(1,n_s)$}
We compute here the pdf $P_T(t)$ of the time $T(1,n_s)$ a single TF takes to bind one of the $n_{s}$ DNA specific targets. Decomposing the pdf
by the event that the target is found after exactly k steps, we have:
\beq
P_T(t)= \sum_{k=0}^{\infty} Pr\{T(1,n_s)<t| k \hbox{ 1D walk} \}
Pr\{ k
\hbox{ 1D walk} \}.
\eeq
Using the probability $p(n_s)$ to bind to one of the $n_s$ sites during a one dimensional motion along the DNA molecule, the probability $\widetilde{P}_{k}=Pr\{ k \hbox{ 1D walk } \}$ to find a site during the $k^{th}$ one dimensional DNA motion is given by:
\beq
\widetilde{P}_{k} &=&p(n_s)(1-p(n_s))^{k-1}.
\label{Pk}
\eeq
A cycle is the concatenation of one and three
dimensional motions. Both periods are characterized by random exponential
 times. The conditional search time for $k$ cycles of DNA binding and free diffusion is then:
\beq
T(1,n_s)|k=\sum\limits_{j=1}\limits^{k}(\tau_{f}(j)+\tau_{d}(j)),
\label{sumtempcondi}\eeq
where $(\tau_{d}(1),..\tau_{d}(k))$ and
$(\tau_{f}(1),..,\tau_{f}(2), ...\tau_{f}(k))$ are respective the times  spent
bound to the DNA and freely diffusing in the nucleus.

To compute $P_T(t)$, we will use the characteristic function $F$ of
$T(1,n_s)$,
\beq
F(x)=\mathbb{E}_{t}(e^{itx})&=&\int\limits_{-\infty}\limits^\infty
e^{itx}p_T(t)dt\\
&=&\sum\limits_{k=1}\limits^\infty G_k(x)\widetilde{P}_k,
\eeq
where $G_k$ is the characteristic function of $(T(1,n_s)<t| k
\hbox{ 1D walk})$. Since the random times $\tau_{d}(j)$ and $\tau_{f}(j)$ are independent,
the characteristic function of the sum (\ref{sumtempcondi}) is the product of the characteristic functions:
\beq
G_k(x)=\prod\limits_{j=1}\limits^k F_{\tau_{f}(j)}(x)
F_{\tau_{d}(j)}(x),
\label{prodcharac}\eeq
where $F_{\tau_{f}(j)}(x)$ and $F_{\tau_{d}(j)}(x)$ are  respectively the characteristic functions of the free diffusion time $\tau_{f}(j)$ and the time $\tau_{d}(j)$ bound to the DNA. Since these times are exponentially distributed:
\beq
F_{\tau_{f}(j)}(x)=\frac{1}{1-ix\overline{\tau}_{f}} \\
F_{\tau_{d}(j)}(x)=\frac{1}{1-ix\overline{\tau}_{d}}.
\eeq
Finally,
\beq
F(x)&=&\sum\limits_{k=1} \limits^\infty
p(n_s)\left(1-p(n_s)\right)^{k-1}\frac{1}{\left(1-ix\overline{\tau}_{f}\right)^k\left(1-ix\overline{\tau}_{d}\right)^k}
\\&=&\frac{p(n_s)}{\left(1-ix\overline{\tau}_{d}\right)\left(1-ix\overline{\tau}_{f}\right)-1+p(n_s)}\label{transflaplace}.
\eeq
The poles are given by the two roots of
$\left(1-y\overline{\tau}_{d}\right)\left(1-y\overline{\tau}_{f}\right)-1+p(n_s)=0$
with $y=ix$:
\beq
r_1=\frac{(\overline{\tau}_{d}+\overline{\tau}_{f})-\sqrt{(\overline{\tau}_{d}+\overline{\tau}_{f})^2-4p(n_s)
\overline{\tau}_{f}\overline{\tau}_{d}}}{2\overline{\tau}_{f}\overline{\tau}_{d}}>0
\label{rootslaplace}
\\r_2=\frac{(\overline{\tau}_{d}+\overline{\tau}_{f})+\sqrt{(\overline{\tau}_{d}+\overline{\tau}_{f})^2-4p(n_s)\overline{\tau}_{f}\overline{\tau}_{d}}}{2\overline{\tau}_{f}\overline{\tau}_{d}}>0,
\nonumber
\eeq
where, for $p(n_s)\in[0;1]$, the two roots $r_1$ and $r_2$ are real
positive.  Decomposing the fraction (\ref{transflaplace}) gives:
\beq
F(x)&=&\frac{p(n_s)}{\overline{\tau}_{d}\overline{\tau}_{f}(r_1-r_2)(ix-r_1)}-\frac{p(n_s)}{\overline{\tau}_{d}\overline{\tau}_{f}(r_1-r_2)(ix-r_2)}
\\&=&\frac{r_1r_2}{(r_1-r_2)(ix-r_1)}-\frac{r_1r_2}{(r_1-r_2)(ix-r_2)},
\eeq
where $\frac{p(n_s)}{\overline{\tau}_{d}\overline{\tau}_{f}}=r_1r_2$ comes from
the equation satisfied  by $r_1$ and $r_2$. 
By inverting the characteristic function
$p_T(t)=\frac{1}{2\pi}\mathbb{E}_{x}(e^{-itx})=\int\limits_{-\infty}\limits^\infty
e^{itx}F(x)dx$ and since the inverse transform of
$-\frac{r_1}{ix-r_1}$ is an exponential distribution of mean
$\frac{1}{r_1}$, we obtain:
\beq
p_T(t)=\frac{r_2}{r_2-r_1}\frac{e^{-tr_1}}{r_1}+\frac{r_1}{r_1-r_2}\frac{e^{-tr_2}}{r_2}.\label{deuxexpos}
\eeq
We conclude that the distribution $p_T$ is the sum of two decreasing
exponentials.
\subsection{Asymptotic pdf of $T(1,n_s)$ for  $p(n_s)\ll 1$}

We shall now study the approximation $p(n_s)\ll 1$, for which:
\beq
r_1&\approx &\frac{p(n_s)}{\overline{\tau}_{d}+\overline{\tau}_{f}}\\
r_2&\approx
&\frac{1}{\overline{\tau}_{d}}+\frac{1}{\overline{\tau}_{f}},
\eeq
and
\beq
p_T(t)\approx\left(1-\eps\right)\frac{p(n_s)}{\overline{\tau}_{d}+\overline{\tau}_{f}}e^{-t\frac{p(n_s)}{\overline{\tau}_{d}+\overline{\tau}_{f}}}+\eps
\,\,
\left(\frac{1}{\overline{\tau}_{d}}+\frac{1}{\overline{\tau}_{f}}\right)e^{-t\left(\frac{1}{\overline{\tau}_{d}}+\frac{1}{\overline{\tau}_{f}}\right)},\eeq
with
$\eps=p(n_s)\frac{1}{(\overline{\tau}_{d}+\overline{\tau}_{f})\left(\frac{1}{\overline{\tau}_{d}}+\frac{1}{\overline{\tau}_{f}}\right)}
\leq\frac{p(n_s)}{4}$. Since $p(n_s)\ll 1$ the second exponential converges faster to $0$ than
the first and is further multiplied by a small coefficient $\eps$.

For a time $ \left(\frac{1}{\overline{\tau}_{d}}+\frac{1}{\overline{\tau}_{f}}\right)t\gg
1$,  we approximate the pdf $p_T$ given in equation (\ref{deuxexpos}) by a single
exponential:
\beq\label{pTsingleexp2}
p_T(t) =
\frac{p(n_s)}{\overline{\tau}_{d}+\overline{\tau}_{f}}e^{-\frac{p(n_s)}{\overline{\tau}_{d}+\overline{\tau}_{f}}t}.
\eeq
Since $\overline{\tau}_{d}$ and $\overline{\tau}_{f}$ are both on the order of a few ms \cite{PLA,Elf}, the single exponential limit is valid for $t$ larger than a few ms.  The mean time $\overline{T}(1,n_{s})$ then reduces to:
\beq\overline{T}(1,n_{s})\approx\frac{\overline{\tau}_{d}+\overline{\tau}_{f}}{p(n_s)}.\eeq

%

\subsection{Computation of $\mathbb{P}_k$ for $\beta\ll1$}

Combining equations (\ref{probantf}), (\ref{decompnsites}) and the first
order approximations in $\beta$, the probability
$\mathbb{P}_{n_{s}}$ that all sites are simultaneously occupied is:
\beq
\mathbb{P}_{n_{s}}&=&e^{-\alpha}\sum\limits_{n_f=n_{s}}\limits^{\infty}\left(1-\frac{\beta n_{s}}{n_f-n_{s}+1}\right)
\frac{ \alpha^{n_f}}{n_f!}.\label{pnsdepart2}
\eeq
Using the partial sum:
\beq
S(x)=\sum\limits_{k=0}\limits^{n_{s}-1}\frac{x^{k}}{k!},
\eeq
and the relations:
\beq
\sum \limits_{n_f=n_s}\limits^{\infty}\frac{\alpha^{n_f}}{n_f!}&=&e^{\alpha}-S(\alpha)\\
\sum\limits_{n_f=n_{s}}\limits^{\infty}\frac{\alpha^{n_f}}{(n_f-n_{s}+1)n_f!}
&=&\alpha^{n_{s}-1}\int\limits_{0}\limits^{\alpha}
\frac{e^{x}-S(x)}{x^{n_{s}}}\, dx,
\eeq
we obtain:
\beq \mathbb{P}_{n_{s}}
&=&1-e^{-\alpha}S(\alpha)-\beta n_{s} e^{-\alpha}\alpha^{n_{s}-1}
\int\limits_{0}\limits^{\alpha}
\frac{e^{x}-S(x)}{x^{n_{s}}} \,dx.\label{alloccupied1}
\eeq
Using the change of variable $x=\alpha u$, we can write:
\beq
\mathbb{P}_{n_{s}}(\alpha)=1-e^{-\alpha}S(\alpha)-\beta n_{s}e^{-\alpha} \int\limits_{0}\limits^{1}
\frac{e^{\alpha u}-S(\alpha u)}{u^{n_{s}}} \,du.\label{alloccupied2}
\eeq
We shall now examine some properties of $\mathbb{P}_{n_{s}}$. For
$\alpha u\geq 0$, $e^{\alpha u}-S(\alpha u)\geq 0$ , thus
$\mathbb{P}_{n_{s}}$ is a decreasing function of $\beta$. Indeed the
partial derivative of $\mathbb{P}_{n_{s}}$ in $\beta$ is negative.
Moreover, $\mathbb{P}_{n_{s}}$ is an increasing function of $\alpha$
for $\beta\ll 1$: starting from expressions (\ref{pnsdepart2}) (which is equal to
(\ref{alloccupied2})) and differentiating with respect to $\alpha$:
\beq
\frac{\p
\mathbb{P}_{n_{s}}}{\p
\alpha}&=&e^{-\alpha}\sum\limits_{n_f=n_{s}}\limits^{\infty}\left(1-\frac{\beta
n_{s}}{n_f-n_{s}+1}\right)
\left(\frac{ \alpha^{n_f-1}}{(n_f-1)!}-\frac{
\alpha^{n_f}}{n_f!}\right).\eeq
Using $n_f\geq n_s$ and $\eps$ sufficiently small, then for $\beta <
\frac{1}{n_s}(1-\eps)$, $\left(1-\frac{\beta
n_{s}}{n_f-n_{s}+1}\right)>\eps$ and we obtain:
\beq
\frac{\p
\mathbb{P}_{n_{s}}}{\p
\alpha}&>&\eps e^{-\alpha} \frac{
\alpha^{n_s-1}}{(n_s-1)!}>0.
\eeq
and $\p
\mathbb{P}_{n_{s}}$ is an increasing function of $\alpha$.

We now proceed with estimating $\mathbb{P}_{n_{s}-1}$. Using equations
(\ref{probantf}) and (\ref{decompnsites}), we obtain:
\beq
\mathbb{P}_{n_{s}-1}&=&e^{-\alpha}\sum\limits_{n_f=n_{s}-1}\limits^{\infty}\mathbb{P}(k=n_s-1|n_f)
\frac{ \alpha^{n_f}}{n_f!}.
\eeq
For $\beta\ll1$, using approximation (\ref{firstorder2}) for the term
in $n_s-1$ and (\ref{firstorder1}) for the other terms:
\beq
\mathbb{P}_{n_{s}-1}&=&e^{-\alpha}\frac{\alpha^{n_{s}-1}}{(n_{s}-1)!}\left(1-\beta\frac{n_{s}-1}{2}\right)+
e^{-\alpha}\sum\limits_{n_f=n_{s}}\limits^{\infty}\frac{\beta
n_{s}}{n_f-n_{s}+1}
\frac{ \alpha^{n_f}}{n_f!}.
\eeq
Using relation (\ref{alloccupied2}), for $\beta \ll 1$, we obtain:
\beq
\mathbb{P}_{n_{s}-1}=e^{-\alpha}\frac{\alpha^{n_{s}-1}}{(n_{s}-1)!}\left(1-\beta\frac{n_{s}-1}{2}\right)+\beta n_{s}e^{-\alpha} \int\limits_{0}\limits^{1}
\frac{e^{\alpha u}-S(\alpha u)}{u^{n_{s}}} \,du.
\label{allbutoneoccupied}
\eeq

Finally, when $k\leq n_s-2$ sites are occupied,{using the first
order approximations for $\mathbb{P}(k |n_f)$ in formula
(\ref{firstorder2}}), we shall only retain the probabilities
associated with $k$ or $k+1$ TFs in the nucleus,
\beq
\mathbb{P}_{k}&\approx&\mathbb{P}(k |k)\mathbb{P}(k)+\mathbb{P}(k |k+1)\mathbb{P}(k+1)\nonumber
\\\nonumber \\&=&\left(1-\frac{k\beta}{(n_{s}-k+1)}\right)
e^{-\alpha}\frac{\alpha^{k}}{k!}+
\frac{(k+1)\beta}{(n_{s}-k)}e^{-\alpha}\frac{\alpha^{k+1}}{(k+1)!}
\nonumber\\
&=&e^{-\alpha}\frac{\alpha^{k}}{k!}\left( 1+\beta\left(\frac{\alpha
}{n_{s}-k}-\frac{k}{n_{s}-k+1} \right) \right).
\label{middleoccuppied}
\eeq


\subsection{Critical value for bistability}

To compute the critical value $R_c$ for which the profile
$\alpha_h(x)$ can be bistable, we use equation:
\beq
\alpha_h=R(1-(1-P(x))(1-f(\alpha_h)))\label{erty-b}.
\eeq
For the critical value $R_c$, the function:
\beq
\alpha\rightarrow R_c(1-(1-P(x))(1-f(\alpha))),\label{functionofalpha}
\eeq
is tangent to $\alpha\rightarrow \alpha$ in $\alpha_c$ for some value of $P(x)$, where
$\alpha_c$ is the point where $f$ changes concavity (figure
\ref{boundary}).  For $k=n_s>1$ and
$\beta=0$,
\beq
f(\alpha)=\mathbb{P}_{n_s}=1-e^{-\alpha}S(\alpha).
\eeq
$f''(\alpha_c)=0$ is equivalent to $S(\alpha_c)-2S'(\alpha_c)+S''(\alpha_c)=0$, where:
\beq
S(\alpha) =\sum\limits_{k=0}\limits^{n_{s}-1}\frac{\alpha^{k}}{k!}.
\eeq
After some computations, we find that:
\beq
\alpha_c=n_s-1.\label{alphacexpression}
\eeq
Now at the critical value $R_c$, the function (\ref{functionofalpha}) is tangent to $\alpha$ in $\alpha_c$ (see figure \ref{boundary}) and
we obtain the conditions:
\beq
R_c(1-(1-P(x))(1-f(\alpha_c)))=\alpha_c\\
R_c\left(1-(1-P(x))\left(1-\frac{\p f}{\p \alpha}(\alpha_c)\right)\right)=1.
\eeq
After simplification,
\beq
1-\frac{1-f(\alpha_c)}{R_c\frac{\p f}{\p \alpha}(\alpha_c)}=\alpha_c/R_c.
\eeq
We then obtain for $R_c$:
\beq
R_c&=&\alpha_c+\frac{1-f(\alpha_c)}{\frac{\p f}{\p \alpha}(\alpha_c)}=\alpha_c+\frac{S(\alpha_c)}{S(\alpha_c)-S'(\alpha_c)}.
\eeq
Finally, using (\ref{alphacexpression}) we obtain:
\beq
R_c&=&n_s-1+(n_s-1)!\frac{S(n_s-1)}{(n_s-1)^{n_s-1}}.\label{Rcritical}
\eeq
and for $n_s=2$, $R_c=3$.

{\noindent {\bf Acknowledgment:} we thank K. Moya for critical
reading of the manuscript.}



\begin{thebibliography}{99}

\bibitem{Firestein} S. Lomvardas, G. Barnea, D. J. Pisapia, M. Mendelsohn,  J. Kirkland, R. Axel,
Interchromosomal interactions and olfactory receptor choice, Cell. 126(2): 403-13 (2006)

\bibitem{VonHippel} O. G. Berg, P. H. von Hippel, Selection of DNA binding sites by regulatory proteins, Trends in Biochemical Sciences. 13(6): 207-211 (1988)

\bibitem{Berg} O. G. Berg, R. B. Winter, P. H. von Hippel, Diffusion-driven mechanisms of protein translocation on nucleic acids. Part 1: Models and theory, Biochemistry. 20(24): 6929-6948 (1981)

\bibitem{blomberg} O. G. Berg, C. Blomberg, Association kinetics with coupled diffusion. An extension to coiled-chain macromolecules applied to the lac repressor-operator system, Biophys Chem. 7(1): 33-9 (1977)

\bibitem{Slutsky} M. Slutsky, L. A. Mirny, Kinetics of protein-DNA interaction: facilitated target location in sequence-dependent potential, Biophys. Journal. 87(6): 4021-35
(2004)

\bibitem{Wunderlich} Z. Wunderlich, L. A. Mirny, Spatial effects on the speed and reliability of protein-DNA search, Nucleic Acids Res. 36(11): 3570-8 (2008)

\bibitem{Halford} S. E. Halford, J. F. Mark, How do site-specific DNA-binding proteins find their targets? Nucleic Acids Research. 32(10): 3040-3052 (2004)

\bibitem{Hu} T. Hu, A. Grosberg, B. Shklovskii, How proteins search for their specific sites on DNA: the role of DNA conformation, Biophys J. 90(8): 2731-44  (2006)

\bibitem{Benichou}  M. Coppey, O. B\'enichou, R. Voituriez, M. Moreau, Kinetics of Target Site Localization of a Protein on DNA: A Stochastic Approach,
Biophys. Journal. 87(3): 1640-9 (2004)

\bibitem{PLA} G. Malherbe, D. Holcman, Search for a DNA target site in the nucleus, PLA. 374(3): 466-471  (2010)

\bibitem{turing} A. M. Turing, The Chemical Basis of Morphogenesis, Phi. Tr. of the Royal Society of London. Series B, Biological Sciences, Vol. 237, No. 641.pp.37-72.(1952)

\bibitem{wolpert}L. Wolpert, One hundred years of positional information,
Trends Genet. 12(9): 359-64 (1996)

\bibitem{meinhart}H. Meinhardt, Models for the generation and interpretation of gradients. Cold Spring Harb Perspect Biol. 1(4):a001362. (2009)

\bibitem{monk1} N. A. Monk, Cell communities and robustness in development, Bull Math Biol. 59(6):1183-9 (1997)



\bibitem{monk2} D. J. Irons, A. Wojcinski, B. Glise, N. A. Monk NA, Robustness of positional specification by the Hedgehog morphogen gradient, Dev Biol. 342(2):180-93  (2010)

\bibitem{Feng} F. He, Y. Wen, J. Deng, X. Lin, L. J. Lu, R.
Jiao, J. Ma, Probing intrinsic properties of a robust morphogen gradient in Drosophila, Dev Cell. 15(4): 558–567 (2008)

\bibitem{Ashyraliyev} M. Ashyraliyev, K. Siggens, H. Janssens, J. Blom, M. Akam, J. Jaeger, Gene circuit analysis of the terminal gap gene huckebein, PLoS Comput Biol. 5(10):e1000548 (2009)

\bibitem{Nadassy} K. Nadassy, S. J. Wodak, J. Janin, Structural features of protein-nucleic acid recognition sites, Biochemistry. 38(7): 1999-2017 (1999)

\bibitem{O'Gorman} R. B. O'Gorman, M. Dunaway, K. S. Matthews, DNA Binding Characteristics of Lactose Repressor and the Trypsinresistant
Core Repressor, Journal
 Bio. Chem. 255(21): 10100-6 (1980)

\bibitem{SchussBook} Z. Schuss, \textit{Theory and applications of
 stochastic differntial equations}. John Wiley Sons Inc, (1980)

\bibitem{Elf}J. Elf, G. W. Li, X. S. Xie, Probing transcription factor dynamics at the single-molecule level in a living cell, Science.  316(5828): 1191-4 (2007)

\bibitem{PNAS} Z. Schuss, A. Singer, D. Holcman, The narrow escape problem for diffusion in cellular microdomains, PNAS. 104(41): 16098-103 (2007)



\bibitem{HS-hole1} D. Holcman, Schuss Z, Diffusion through a cluster of small windows and flux regulation in microdomains, Physics Letters A. 372(21): 3768-3772 (2008)

\bibitem{HS-hole2} D. Holcman, Schuss Z, Diffusion escape through a cluster of small absorbing windows,
Journal of Physics A. 41: 155001 (2008)

\bibitem{WardM1} S. Pillay,  M .Ward, A. Peirce, T. Kolokolnikov, An Asymptotic Analysis of the Mean First Passage Time for Narrow Escape Problems: Part I: Two-Dimensional Domains, SIAM Multiscale Modeling and
Simulation. 8(3): 803-835 (2010)

\bibitem{WardM2}A. Cheviakov, M. Ward, R. Straube, An Asymptotic Analysis of the Mean First Passage Time for Narrow Escape Problems: Part II: The Sphere, SIAM Multiscale Modeling and Simulation,   8(3): 836-870 (2010)

\bibitem{WardM3} A. Cheviakov, M .Ward, Optimizing the Principal Eigenvalue of the
Laplacian in a Sphere with Interior Traps, Mathematical and Computer
Modeling, In press, (2010)

\bibitem{Berg-Purcell}H. C. Berg, E.M. Purcell, Physics of chemoreception,
Biophys J. 20(2):193-219 (1977)

\bibitem{zwanzig}  R. Zwanzig, Diffusion-controlled ligand binding to spheres partially covered by receptors: an effective medium treatment, Proc.
Nat. Acad. Sci. 87(15):5856-5857 (1990).

\bibitem{bere}  A. M. Berezhkovskii, Y. A. Makhnovskii, M. I. Monine, V. Y. Zitserman,
S. Y. Shvartsman, Boundary homogenization for trapping by patchy surfaces, J Chem Phys. 121(22):11390-4 (2004)

\bibitem{Baliga} R. Baliga, E. E. Baird, D. M. Herman, C. Melander, P. B. Dervan, D. M. Crothers, Kinetic Consequences of Covalent Linkage of DNA Binding Polyamides,
Biochemistry, 40(1):3-8 (2001)

\bibitem{Lin}S. Y. Lin, A. D. Riggs, Lac operator analogues: bromodeoxyuridine substitution in the lac operator affects the rate of dissociation of the lac repressor, PNAS. 69(9): 2574-76 (1972)

\bibitem{HolcmanJCP2005} D. Holcman, Z. Schuss, Stochastic chemical reactions in microdomains, Journal of Chemical Physics. 122(11): 114710 (2005)

\bibitem{Ptaschnebook} M. Ptashne, \textit{A genetic switch: phage lambda revisited}. Cold Spring Harbor Laboratory Press (2004)

\bibitem{Wang} L. Wang, B. L. Walker, S. Iannaccone, D. Bhatt, P. J. Kennedy, W. T. Tse, Bistable switches control memory and plasticity in cellular differentiation,
PNAS. 106(16): 6638-6643 (2009)

\bibitem{Crews}S. T. Crews, J. C. Pearson, Transcriptional autoregulation in development, Curr Biol. 19(6): 241-246 (2009)

\bibitem{Lopes}J. P. Lopes, F. M. Vieira, D. M. Holloway, P. M. Bisch, A. V. Spirov,
Spatial Bistability Generates hunchback Expression Sharpness in the
Drosophila Embryo, PLoS Comput Biol. 4(9):e1000184.

\bibitem{Struhl}G. Struhl, K. Struhl, P. M. Macdonald, The gradient morphogen bicoid is a concentration-dependent transcriptional activator, Cell. 57: 1259–1273 (1989)

\bibitem{Driever} W. Driever, C. Nusslein-Volhard, The bicoid protein is a positive regulator of hunchback transcription in the early Drosophila embryo, Nature. 337: 138–143 (1989)

\bibitem{Burz} D. S. Burz, R. Rivera-Pomar, H. Jackle, S. D. Hanes, Cooperative
DNA-binding by Bicoid provides a mechanism for threshold-dependent
gene activation in the Drosophila embryo, EMBO J. 17(20):5998-6009
(1998)

\bibitem{Pisarev} A. Pisarev, E. Poustelnikova, M. Samsonova, J. Reinitz, FlyEx, the quantitative atlas on segmentation gene expression at cellular resolution, Nucl. Acids Res. 37: 560-566.(2009)

\bibitem{Poustelnikova} E. Poustelnikova, A. Pisarev, M. Blagov, M. Samsonova, J. Reinitz, A database for management of gene expression data in situ, Bioinformatics. 20: 2212-2221 (2004)

\bibitem{Lebrecht} D. Lebrecht, M. Foehr, E. Smith, F. J. P.Lopes, C. E. Vanario-Alonso, J. Reinitz,
D. S. Burz, S. D. Haneet, Bicoid cooperative DNA binding is critical for embryonic patterning in
Drosophila. PNAS 102: 13176–13181.

\bibitem{Ptashne2} M. Ptashne, A. Gann \textit{Genes and Signals} Cold Spring Harbor Laboratory Press, (2002).

\bibitem{Gregorlimits} T. Gregor, D. W. Tank, E. F. Wieschaus, W.  Bialek, Probing the limits to positional information, Cell. 130(1):153-64 (2007)

\bibitem{Lohr}U Lohr, H. R. Chung, M. Beller, H. Jackle, Antagonistic action of Bicoid and the repressor
Capicua determines the spatial limits of Drosophila
head gene expression domains, PNAS. 106(51):21695-700. (2009)

\bibitem{Wu} X. Wu, R. Vakani, S. Small, Two distinct mechanisms for differential positioning of gene expression borders involving the Drosophila gap protein giant, Development. 125(19):3765-74. (1998)

\bibitem{Kasatkin}V. Kasatkin, A. Prochiantz, D.  Holcman,
Bull Math Biol, Morphogenetic gradients and the stability of boundaries between neighboring morphogenetic, 70(1):156-78 (2008)

\bibitem{DHolcmangradient}D. Holcman, V. Kasatkin, A. Prochiantz, Modeling homeoprotein intercellular transfer unveils a parsimonious mechanism for gradient and boundary formation in early brain development, J Theor Biol. 249 (3): 503-17. (2007)

\end{thebibliography}
\end{document}